\documentclass[11pt]{article}

\usepackage[final]{acl}

\usepackage{times}
\usepackage{latexsym}

\usepackage[T1]{fontenc}
\usepackage[utf8]{inputenc}

\usepackage{microtype}

\usepackage{inconsolata}

\usepackage{graphicx}

\usepackage{enumitem}
\usepackage{booktabs}
\usepackage{makecell}
\usepackage{siunitx} 
\usepackage{multirow}
\usepackage{caption} 
\usepackage{etoolbox}
\usepackage{subcaption} % For creating sub-tables (side-by-side)
\usepackage{array}    % Required for the m{} column type for vertical centering
\usepackage{threeparttable} % For table notes
\usepackage{listings}
\usepackage{xcolor} % For adding color
\usepackage{enumitem}
\usepackage{longtable}
\usepackage{supertabular}
\usepackage{booktabs}
\usepackage{enumitem}
\usepackage{url}

\usepackage{breakurl}
\usepackage[breaklinks]{hyperref}
\usepackage{afterpage} 
\usepackage[compact]{titlesec}

\usepackage{cuted} % Add this package for the 'strip' environment

\usepackage{enumitem} % For custom list formatting (e.g., "Step 1:")
\usepackage{xcolor}   % For defining custom colors
\usepackage{listings} % For formatting code blocks
\usepackage[most]{tcolorbox} % For creating the colored boxes

\definecolor{codegray}{gray}{0.95}
\definecolor{codegray-border}{gray}{0.85}
\definecolor{boxgreen}{RGB}{230, 245, 230}
\definecolor{boxred}{RGB}{255, 235, 235}
\definecolor{bordergreen}{RGB}{80, 150, 80}
\definecolor{borderred}{RGB}{180, 80, 80}
\definecolor{titlegreen}{RGB}{0, 100, 0}
\definecolor{titlered}{RGB}{139, 0, 0}

\lstdefinestyle{mystyle}{
    backgroundcolor=\color{codegray},
    commentstyle=\color{green!40!black},
    keywordstyle=\color{blue},
    numberstyle=\tiny\color{gray},
    stringstyle=\color{purple},
    basicstyle=\ttfamily\small,
    breakatwhitespace=false,
    breaklines=true,
    captionpos=b,
    keepspaces=true,
    numbers=none,
    numbersep=5pt,
    showspaces=false,
    showstringspaces=false,
    showtabs=false,
    tabsize=2,
    frame=single,
    rulecolor=\color{codegray-border},
    frameround=tttt,
    xleftmargin=4pt,
    xrightmargin=4pt,
    literate={/}{/}{1\discretionary{}{}{}}
             {.}{.}{1\discretionary{}{}{}}
             {=}{=}{1\discretionary{}{}{}},
}
\newtcolorbox{successbox}[1][]{
  breakable, 
  enhanced,
  colback=boxgreen,
  colframe=bordergreen,
  boxrule=1pt,
  fonttitle=\bfseries,
  coltitle=titlegreen,
  title=#1,
  attach boxed title to top left={yshift=-0.25cm, xshift=0.5cm},
  boxed title style={
    colback=white,
    colframe=bordergreen,
    boxrule=0.5pt,
    arc=3mm
  },
  top=12pt,
  bottom=8pt,
  left=6pt,
  right=6pt
}

\newtcolorbox{failurebox}[1][]{
  breakable,
  enhanced,
  colback=boxred,
  colframe=borderred,
  boxrule=1pt,
  fonttitle=\bfseries,
  coltitle=titlered,
  title=#1,
  attach boxed title to top left={yshift=-0.25cm, xshift=0.5cm},
  boxed title style={
    colback=white,
    colframe=borderred,
    boxrule=0.5pt,
    arc=3mm
  },
  top=12pt,
  bottom=8pt,
  left=6pt,
  right=6pt
}

\title{Automating Android Build Repair: Bridging the Reasoning-Execution Gap in LLM Agents with Domain-Specific Tools}

\author{Ha Min Son\textsuperscript{1,2} 
 Huan Ren\textsuperscript{2} 
 Xin Liu\textsuperscript{1} 
 Zhe Zhao\textsuperscript{1,*} \\
\textsuperscript{1}Department of Computer Science, University of California, Davis \\
\textsuperscript{2}CodeDroid LLC \\
}

\begin{document}
\maketitle
\begin{abstract}
Android is the largest mobile platform, yet automatically building applications remains a practical challenge. While Large Language Models (LLMs) show promise for code repair, their use for fixing Android build errors remains underexplored. To address this gap, we first introduce AndroidBuildBench, a benchmark of 1,019 build failures curated from the commit histories of 43 open-source Android projects. Each problem is paired with a verified solution from a subsequent commit, ensuring that fixes are feasible. Second, we propose GradleFixer, an LLM agent with domain-specific tools for inspecting and manipulating the Gradle build environment. GradleFixer achieves a resolve rate of 81.4\% (pass@1), significantly outperforming a state-of-the-art coding agent that relies on a general-purpose shell. 
GradleFixer's success suggests that while LLMs possess the high-level knowledge to solve these failures, they struggle to translate this knowledge into effective low-level actions using a general-purpose shell.
We demonstrate the effectiveness of a strategy we term \textit{Tool Bridging}, which replaces general-purpose shell commands with domain-aware abstractions. We hypothesize this approach works through two mechanisms: 1) it provides tools in an API-like format that LLMs use more reliably, and 2) it constrains the action space to relevant operations. This approach bridges the gap between the model's high-level reasoning and effective low-level execution.
\end{abstract}

\section{Introduction}

\begin{figure}[htbp]
    \centering 
    \hspace{-20pt}
    \vspace{-20pt}
    \includegraphics[width=1\columnwidth]{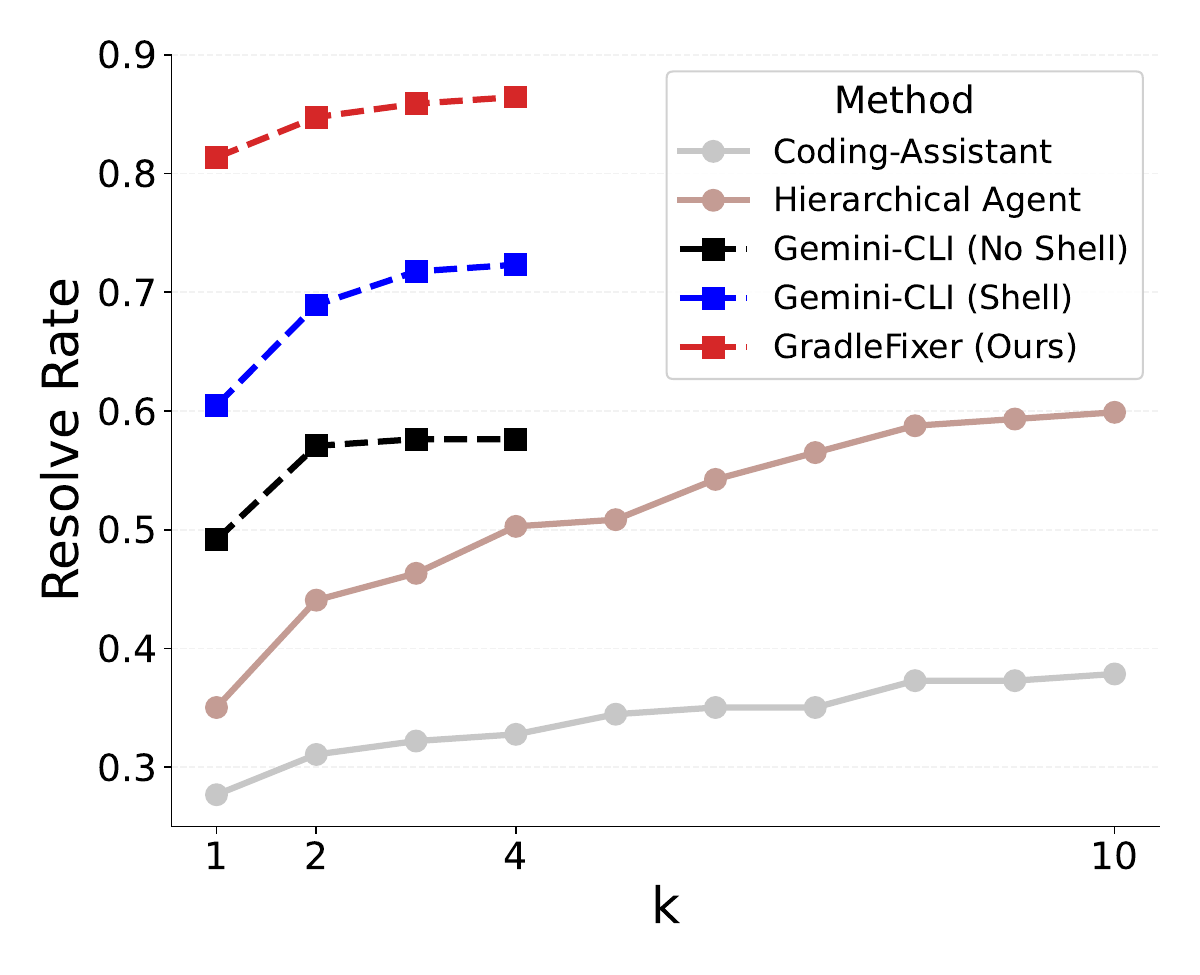}
    \caption{The pass@k resolve rates (percentage of problems solved within k independent sampling attempts) for different agent frameworks on our test set of 184 build errors. We find that replacing the general \texttt{shell} with domain-specific tools significantly improves performance. }
    \label{fig:main_results}
    \vspace{-15pt}
\end{figure}

Android is the largest and most active mobile development platform in the world, with a market share of 73.9\% of the global mobile OS market, over 2.2 million apps on the Google Play Store, and over 1.4 million questions tagged "android" on Stack Overflow \citep{statista_googleplay_2024, statcounter_os_share, stackoverflow_android_questions}. Despite its popularity, a practical challenge remains: \textit{applications cannot reliably be built}. A comprehensive study of 5,222 apps revealed that only 31.3\% could be built successfully out-of-the-box \citep{liu2024understanding}. In our dataset, we find the primary source of issues is syntax errors (Table \ref{tab:root_cause_sub}). While Continuous Integration and Continuous Deployment (CI/CD) pipelines are a well-established practice for mitigating these errors \citep{rostami2023usage, baitha2024streamlining, ghaleb2024ci, liu2022first}, \citet{liu2024understanding} reported only 21.8\% out of the 5,222 apps had CI/CD scripts. Furthermore, even with CI/CD, maintaining a stable build requires significant effort. \citet{hassan2017automatic} found that over 74\% of CI/CD configuration changes are dedicated to build fixing and environment updates. This maintenance is necessary to address a complex range of errors, including syntax mistakes, missing resource files, configuration errors, and library unavailability. 

Large Language Models (LLMs) and their application in Automated Program Repair (APR) offer a promising alternative \citep{zhang2024systematic, yin2024thinkrepair, xia2023automated, jin2023inferfix}. However, the application of LLMs to Android build errors remains an underexplored area. Existing LLM coding benchmarks, like SWE-bench \citep{jimenez2023swe}, primarily evaluate models on resolving bugs from GitHub issues, not the transient, environment-specific build failures common in Android development. Furthermore, existing datasets of Android build failures~\citep{liu2024understanding, hassan2017automatic} capture a repository at a single, broken point in time, making fixes difficult to verify as feasible without a corresponding working build to serve as ground truth. 

To address these limitations, we make two primary contributions. First, we introduce AndroidBuildBench, a benchmark of real-world Android build errors. AndroidBuildBench contains 1,019 reproducible build failures curated from the commit histories of 43 popular open-source Android projects. Each failure is paired with a verified solution from a subsequent commit, ensuring all problems are feasible to fix. In addition to common human errors and dependency conflicts, our benchmark includes a category of LLM-generated errors to reflect the changing nature of software development.

Second, we propose \textbf{GradleFixer}, an agent-based approach for build error fixing using domain-specific tools. Inspired by prior work showing that specialized tools improve agent performance \citep{singh2025code, wang2023voyager} (details in Sec.  \ref{app:relation_to_prior_work}), GradleFixer equips an LLM agent with domain-specific tools, which are simple wrappers for shell commands, to inspect and manipulate the Gradle build environment. Our experiments show that GradleFixer significantly outperforms state-of-the-art baselines across all categories of build failures (Figure \ref{fig:main_results}). 
We infer from GradleFixer's success that LLMs have the high-level knowledge needed to fix build errors. The poorer performance of shell-based agents, however, suggests this knowledge fails to transfer to effective low-level execution. This gap explains why other effective agents also rely on specialized toolsets \citep{wang2023voyager, singh2025code}, although these studies do not explore this choice as a research question. Our approach, which we term \textit{Tool Bridging}, addresses this gap by replacing a general-purpose shell with domain-aware abstractions to bridge high-level reasoning to correct low-level execution.
We hypothesize this strategy improves performance through two simultaneous mechanisms. First, by providing API-like tools that abstract away complex command syntax, it allows the model to focus on reasoning rather than implementation. Second, it constrains the action space, reducing irrelevant operations. By reframing the task into a format the model uses more reliably~\citep{liu2024toolace, zhang2024ecoact, zhang2025nemotron}, our strategy connects the LLM's high-level reasoning to effective low-level execution.
Our contributions are summarized as follows:
\begin{itemize}[nosep, leftmargin=1.5em]
\item We introduce \textbf{AndroidBuildBench}, a benchmark of 1,019 reproducible Android build errors curated from commit histories. By pairing failing commits with their subsequent fixes, we ensure problems are verifiable and solvable.
\item We introduce \textbf{GradleFixer} and demonstrate that our strategy of \textit{Tool Bridging}, which replaces general tools with domain-specific abstractions, is highly effective for Android build repair. We hypothesize this approach connects high-level reasoning to correct low-level execution by providing API-like abstractions and constraining the action space.
\end{itemize}

\begin{table}[t!]
  \centering
  \small
  \caption{Summary statistics of the 43 curated Android projects used in AndroidBuildBench.}
  \label{tab:project_stats}
  \begin{tabular}{lrr}
    \toprule
    \textbf{Metric} & \textbf{Mean ± SD} & \textbf{Median} \\
    \midrule
    Stars                   & 3,887 $\pm$ 4,702   & 1,500   \\
    Pull Requests           & 866 $\pm$ 1,429    & 251      \\
    Commits                 & 4,258 $\pm$ 6,807   & 1,263    \\
    Files                   & 1,041 $\pm$ 1,198   & 646      \\
    Lines of Code           & 117,971 $\pm$ 109,442 & 94,699  \\
    Last Commit (Days)      & 138  $\pm$ 332    & 14         \\
    \bottomrule
  \end{tabular}
  \vspace{-20pt}
\end{table}

% \begin{figure}[htbp]
%     \centering 
%     \includegraphics[width=\columnwidth]{latex/images/data_curation.pdf}
%     \caption{ }
%     \label{fig:data_curation}
% \end{figure}

\section{AndroidBuildBench: A Benchmark for Android Build Repair}

\subsection{Project Curation}
AndroidBuildBench is curated from 43 active Android projects on GitHub. These were selected by filtering the top 100 most-starred applications for Java/Kotlin usage, >500 stars, >100 closed pull requests, recent commits, and a Gradle build script. This process yielded 43 open-source projects (Appendix \ref{app:project_list}). A summary of the project statistics is shown in Table \ref{tab:project_stats}.
We focus on the Gradle build system, as it is used by 95.1\% of Android applications according to a recent study \citep{liu2024understanding}.

Our approach focuses on individual commits. We are able to explore the specific changes that introduce build failures and the subsequent commits that fix them, which allows us to create problem-solution pairs. Unlike prior work that analyzes repositories at a single point in time without providing solutions, our approach creates verifiable problem-solution pairs.

\begin{table}[htbp]
    \centering
    \small
    \caption{Distribution of errors in the test set compared to the entire dataset.}
    \label{tab:dataset_distribution}
    \begin{tabular}{l S[table-format=2] S[table-format=3]}
        \toprule
        \textbf{Error Category} & {\textbf{Full Dataset}} & {\textbf{Test Set}} \\
        \midrule
        Human-Committed     & 341 & 66  \\
        Augmented Dependency& 486 & 43  \\
        LLM-Generated       & 192 & 75  \\
        \midrule
        \textbf{Total Instances} & \textbf{1019} & \textbf{184}  \\
        \midrule
        Contributing Projects & 43 & 26 \\
        \bottomrule
    \end{tabular}
    \vspace{-15pt}
\end{table}

\subsection{Build Error Curation}
From the 43 curated projects, we created a dataset of 1,019 reproducible build failures. For fine-grained analysis, we created a test set of 184 instances randomly sampled from 26 projects, ensuring a diverse and representative subset suitable for manual inspection. An overview of the dataset distribution is provided in Table \ref{tab:dataset_distribution}.

The benchmark contains errors from three curation methods designed to capture real-world build challenges, which are detailed below.

\noindent \textbf{Human-Committed Errors.} This category captures the transient build failures that naturally occur during the software development lifecycle. We identify merged pull requests (PR) that successfully build, then analyze their commit history. If an intermediate commit within the PR fails to build, we classify this failing commit as a problem instance. The corresponding solution consists of the changes that lead to the final, successful PR.

\noindent \textbf{Augmented Dependency Errors.} To simulate configuration issues, we start with a successful commit and revert only its build-related file changes (e.g., build.gradle, Manifest.xml). A build failure resulting from this change simulates a common scenario where source code becomes out-of-sync with its build environment or dependency declarations. This process generates a targeted problem instance where the solution is known to be contained within the reverted configuration changes.

\noindent \textbf{LLM-Generated Errors.} To capture failures from AI-assisted development, we revert a successful commit and prompt an LLM with the commit message and the release notes to re-implement the functionality. If the code generated by the LLM fails to build, we include this failure in our benchmark. The successful human-written commit is the ground-truth solution, reflecting a realistic scenario where a developer might need to fix an imperfect, AI-generated patch. This data generation process is also highly scalable, allowing us to bootstrap a large number of diverse LLM-generated failures.

\begin{table*}[t]
    \centering    
    \caption{Analysis of build failures and code changes by origin.}
    \label{tab:combined_analysis}
    \begin{subtable}{0.55\textwidth}
        \centering
        \small
        \caption{Root Cause Analysis of Failures.}
        \label{tab:root_cause_sub}
        \begin{tabular}{l rrr r}
            \toprule
            \textbf{Root Cause}     & \textbf{Human} & \textbf{Dep.} & \textbf{LLM} & \textbf{Total (\%)} \\
            \midrule
            Syntax code             & 35 & 28 & 47 & 110 (59.8\%) \\
            Resource file missing   & 0  & 5  & 7  & 12 (6.5\%) \\
            Configuration error     & 27 & 5  & 14 & 46 (25.0\%) \\
            Library not available   & 4  & 5  & 7  & 16 (8.7\%) \\
            NDK error               & 0  & 0  & 0  & 0 (0.0\%) \\
            \midrule
            \textbf{Total}          & \textbf{66} & \textbf{43} & \textbf{75} & \textbf{184 (100\%)} \\
            \bottomrule
        \end{tabular}
    \end{subtable}%
    \hfill 
    \begin{subtable}{0.42\textwidth}
        \centering
        \small
        \caption{Statistics of Code Changes.}
        \label{tab:change_metrics_sub}
        \begin{tabular}{ll S[table-format=3.1] S[table-format=2.1] S[table-format=3.1]}
            \toprule
            \textbf{Metric} & \textbf{Stat.} & {\textbf{Human}} & {\textbf{Dep.}} & {\textbf{LLM}} \\
            \midrule
            \multirow{3}{*}{\shortstack[l]{Files\\Changed}} 
                            & Mean      & 5.8   & 14.3  & 6.7   \\
                            & Median    & 3.0   & 14.0  & 4.0   \\
                            & Std Dev.  & 4.8   & 9.4   & 7.0   \\
            \midrule
            \multirow{3}{*}{\shortstack[l]{Lines\\Changed}} 
                            & Mean      & 258.6 & 436.9 & 274.1 \\
                            & Median    & 66.0  & 229.0 & 81.0  \\
                            & Std Dev.  & 467.0 & 519.1 & 457.9 \\
            \bottomrule
        \end{tabular}
    \end{subtable}
    \vspace{-10pt}
\end{table*}

\subsection{Error Categorization}
We manually analyzed the 184 test set issues, adopting the five build failure categories proposed by \citet{liu2024understanding}.
\begin{itemize}[nosep, leftmargin=1em]
    \item \textbf{Syntax Error:} errors within the Java or Kotlin source code. These are fundamental language mistakes, such as a missing semicolon or an incorrect keyword, that cause compiler errors.
    \item \textbf{Resource File Missing:} errors when the build process cannot find a required file. These files, such as keystore.properties or google-services.json, may be missing unintentionally or deliberately excluded from version control to protect confidential information.
    \item \textbf{Configuration Error:} errors from incorrect settings in project configuration files, most commonly build.gradle. A typical cause is a hardcoded, environment-specific path (e.g., to the Android SDK) that does not exist on the current build machine.
    \item \textbf{Library Not Available:} errors when the build system is unable to locate and download a specified dependency library. This can happen if a library has been removed from its repository, the repository URL is invalid, or the specified version is obsolete.
    \item \textbf{NDK Error:} errors in projects that use the Android Native Development Kit (NDK) to include C/C++ code. Errors often involve misconfigurations, such as a mismatch between the NDK version required by the project and the one installed, or build script issues related to changes in the NDK package structure over time.
\end{itemize}

Our analyses are presented in Table \ref{tab:root_cause_sub}. A notable finding is the absence of NDK errors. We posit this absence results from our curation method, which is anchored to successfully built pull requests. NDK errors often result from build machine environment incompatibilities. Since our method requires a successful build as an endpoint, it naturally filters out states with systemic environment issues. The errors captured are those transiently introduced and fixed within a pull request, a scope more representative of the typical software development life cycle.
Furthermore, we find that syntax errors are the most common failure type. This is surprisingly high for human-committed errors, given that modern IDEs typically provide real-time syntax checking. For LLM-generated errors, this result is less surprising, as current models are known to be prone to syntactical mistakes \citep{tambon2025bugs}. 

Furthermore, we analyzed the statistical properties of the code changes, as shown in Table \ref{tab:change_metrics_sub}. We find that Augmented Dependency Errors involve substantially larger code changes. Our curation method for these errors selects commits with build file modifications, which often correlate with more substantial source code changes. As we demonstrate in our results (Table \ref{tab:success_failure_analysis}), the magnitude of the code change (i.e., files and lines modified) that led to the error is a reasonable proxy for the difficulty of the repair task, rather than the category of the error. A full distribution is in Appendix \ref{appendix:problem_diff}.

\section{Experimental Setup and Approaches}
\label{sec:setup}

\subsection{Motivation: Providing the Right Tools}
LLMs demonstrate strong proficiency with common shell commands and can often infer how to use novel tools from their descriptions \citep{patil2025bfcl, zeng2025glm, tbench_2025}. However, our preliminary analysis revealed a critical limitation when applying this general capability to a specialized domain like Android build repair. Although an LLM may possess the high-level knowledge to understand a build failure, it struggles to translate this knowledge into the correct sequence of low-level shell commands. For example, correctly synthesizing and sequencing a complex command like \texttt{./gradlew assembleDebug} along with necessary environment variables is highly error-prone compared to the high-level decision of when to run a build.

We find this gap between high-level reasoning and low-level execution (Tables \ref{tab:main_results_wide} \& \ref{tab:tool_usage_counts}) likely occurs because while common shell commands (e.g., \texttt{ls}, \texttt{grep}) are prevalent in training data \citep{liu2024toolace, patil2024gorilla}, the multi-step sequences required for domains like Android builds are less common. An agent with a general shell often knows the relevant commands individually but fails to apply them in the correct context, leading to inefficient trial-and-error loops. This phenomenon is analogous to providing a specialist with a curated set of instruments rather than an exhaustive, unorganized toolkit.

Our approach addresses this through \textit{Tool Bridging}, which replaces general-purpose tools with domain-aware abstractions. These specialized tools translate shell commands into high-level, API-like actions, presenting a format that LLMs handle more reliably and constraining the action space~\citep{patil2025bfcl}. This design allows the model to focus its reasoning on debugging rather than command synthesis, bridging the gap between conceptual understanding and correct execution.

\subsection{Execution Environment}

All experiments were conducted on a standardized Linux machine (256 CPU cores, 512 GB RAM) installed with the Android SDK. To accommodate the diverse requirements of different projects, we installed the major versions of the Java Development Kit (JDK), including versions 11, 17, 20, 21, 22, and 23. Following the methodology of prior work~\citep{liu2024understanding}, we build each application using the command \texttt{./gradlew assembleDebug -{}-parallel}. Building Android applications is computationally intensive, with build times for some projects extending up to 15 minutes on a 256-core machine. We containerized each build process within an isolated Jupyter Notebook~\citep{kluyver2016jupyter}. This approach enabled us to run multiple tasks concurrently, significantly improving experimentation throughput. Additionally, to ensure a clean state between runs, we run \texttt{./gradlew clean -{}-stop} before attempting to build, preventing cached states from affecting the current build.

\begin{table*}[th!] 
    \centering
    \small
    \caption{Resolve rate of agent configurations on the AndroidBuildBench test set, measured by Pass@k success rates (\%). Our method, \textbf{GradleFixer}, consistently outperforms all baselines across all failure categories. Best results in each column are in bold. (LLM calls are not budgeted.)}
    \label{tab:main_results_wide}
    \sisetup{table-format=2.1} 
    \begin{tabular}{l SSS SSS SSS}
        \toprule
        & \multicolumn{3}{c}{\textbf{Human Commit}} 
        & \multicolumn{3}{c}{\textbf{Dependency}} 
        & \multicolumn{3}{c}{\textbf{LLM-Generated}} \\
        \cmidrule(lr){2-4} \cmidrule(lr){5-7} \cmidrule(lr){8-10}
        \textbf{Method} & {\textbf{P@1}} & {\textbf{P@2}} & {\textbf{P@4}} 
                        & {\textbf{P@1}} & {\textbf{P@2}} & {\textbf{P@4}} 
                        & {\textbf{P@1}} & {\textbf{P@2}} & {\textbf{P@4}} \\
        \midrule
        Coding-Assistant          & 30.2 & 32.6 & 32.6 & 19.4 & 19.4 & 20.9 & 36.0 & 42.7 & 44.0 \\
        Hierarchical Agent        & 27.9 & 41.9 & 51.2 & 37.9 & 40.9 & 47.0 & 36.0 & 49.3 & 53.3 \\
        Gemini-CLI (No Shell)     & 39.5 & 46.5 & 48.8 & 42.4 & 47.0 & 47.0 & 60.0 & 70.7 & 71.7 \\
        Gemini-CLI (Shell)        & 65.1 & 69.8 & 79.1 & 40.9 & 50.0 & 50.0 & 72.0 & 81.3 & 82.7 \\
        \textbf{GradleFixer (Ours)} & \textbf{84.1} & \textbf{90.9} & \textbf{90.9} 
                                  & \textbf{77.8} & \textbf{79.4} & \textbf{82.5} 
                                  & \textbf{82.3} & \textbf{84.8} & \textbf{85.9} \\
        \bottomrule
    \end{tabular}
    \vspace{-10pt}
\end{table*}

\subsection{Approaches}

We evaluate the effectiveness of an agent-based approach to repairing Android build errors by comparing several configurations designed to isolate the impact of domain-specific tooling. All of our agent configurations use Gemini-2.5-Pro \citep{comanici2025gemini} as the core LLM.

As a baseline representing a state-of-the-art coding agent, we use the official, unmodified Gemini-CLI open-source implementation \citep{google_gemini_cli}. To allow for controlled experiments, we also developed a validated Python replica of the Gemini-CLI agent (Appendix \ref{app:replica_validation}). This replica preserves the original agentic loop and prompting (Appendix \ref{appendix:prompts}) but allows us to easily customize the available toolset. We use this replica for the \textbf{Gemini-CLI (Read/Write Only)} configuration and our proposed method, \textbf{GradleFixer}. We evaluate the following five distinct approaches:

\begin{itemize}[nosep,leftmargin=1em]
    \item \textbf{Coding-Assistant:} A baseline that simulates a user-in-the-loop assistant. The agent is provided with file context and an error log but can only propose code modifications. It cannot execute commands, browse the file system, or run the build. We use the open-source Aider~\citep{aider-website} framework for this configuration.
    \item \textbf{Hierarchical Agent:} A two-agent framework where a primary LLM agent reasons about the problem and then delegates code editing tasks to the \textbf{Coding-Assistant} by invoking it as a tool, similar to \citet{ishibashi2024self, liu2024towards}.
    \item \textbf{Gemini-CLI (Read/Write Only):} A baseline configuration (using our Python replica) where the agent has the ability to read and write files but is denied access to the shell. This configuration tests the agent's ability to solve build errors with a limited toolset.
    \item \textbf{Gemini-CLI (Read/Write and Shell):} The standard open-source Gemini-CLI agent \cite{google_gemini_cli}. \textit{We use the unmodified open-source version for this baseline.} It is given file read/write capabilities and a general-purpose shell tool, allowing it to execute arbitrary commands.
    \item \textbf{GradleFixer (Our Method):} Our proposed method, which excludes the general-purpose shell but is given a set of domain-specific tools designed for the Android build environment.
\end{itemize}

\subsection{GradleFixer Domain-Specific Tools}
The standard tools available to agents, as in Gemini-CLI, include the following file system commands: \texttt{ls}, \texttt{grep}, \texttt{glob}, \texttt{read\_file}, and \texttt{replace\_string}. Agents can also use a \texttt{search\_google} tool, which queries Google and returns summarized findings. Our method adds three domain-specific tools, replacing the general \texttt{shell} tool:

\begin{itemize}[nosep,leftmargin=1em]
    \item \texttt{run\_build}: Executes the standard build (\texttt{./gradlew assembleDebug}) and returns the raw output.
    \item \texttt{run\_gradle}: A more general tool that can execute any Gradle command (\texttt{./gradlew <task>}). This tool is a superset of \texttt{run\_build}.
    \item \texttt{change\_java\_version}: A dedicated function that switches the active JDK in the environment (e.g., \texttt{export JAVA\_HOME=...}).
\end{itemize}

All three tools are intentionally designed as wrappers for specialized shell commands. The outputs of these tools are not specially structured or processed. This design ensures the output is identical to that of a raw shell command, thereby isolating the impact of the tool invocation itself from any benefit of structured output. Full tool descriptions are available in Appendix \ref{appendix:tools}.

\section{Results and Discussions}

The performance of each agent configuration on the AndroidBuildBench test set is shown in Figure \ref{fig:main_results} and Table \ref{tab:main_results_wide}. Our proposed method, \textbf{GradleFixer}, significantly outperforms all baseline configurations across all categories of build errors.

The improvement of Gemini-CLI over \textit{Coding-Assistant} and \textit{Hierarchical Agent} confirms that Android build repair is not a localized code-fixing task. It is a holistic problem requiring an agent to explore the repository, inspect files, and interact with the build system. Additionally, we find that Gemini-CLI (Shell) outperforms \textit{No Shell}. Without a shell, the agent cannot execute Gradle commands, including building the app to inspect the current state, or modify environmental variables like \texttt{JAVA\_HOME}, which are important for resolving errors.

While the open-source Gemini-CLI already abstracts some shell commands (e.g., \texttt{ls}, \texttt{grep}) into dedicated tools, our method takes this direction further. We find that replacing the generic \texttt{shell} with the domain-specific tools \texttt{run\_build}, \texttt{run\_gradle}, and \texttt{change\_java\_version}, leads to a dramatic performance improvement. The most substantial gain was on Dependency issues, which is where Gemini-CLI (Shell) struggled the most.

\subsection{Impact of domain-specific tools}
We study the impact of the domain-specific tools, which are all abstractions of shell commands. We evaluated agents equipped with different tool combinations, from the most general (\texttt{shell}) to the most specific (\texttt{run\_build}). The results are shown in Table \ref{tab:domain_tools}. For ablations, LLMs are budgeted to 30 calls per task.

\begin{table}[t!]
    \centering
    \small
    \caption{Performance of domain-specific tool combinations. The agent's available tools are grouped by category. \textit{(LLMs are limited to \textbf{30 calls} for ablations resulting in slightly lower resolve rates}.)}
    \label{tab:domain_tools}
    \setlength{\tabcolsep}{10pt} % Adjust space between columns
    \begin{tabular}{@{} l c @{}} % @{} removes padding at the edges
        \toprule
        \textbf{Method / Tool Set} & \textbf{Pass@1}  \\
        \midrule
        No shell (Baseline) & 49.2\%  \\
        \midrule
        \multicolumn{2}{@{}l}{\textbf{Individual Tools (w/ Incr. Specificity)}} \\
        \hspace{1em}Only shell      & 54.3\%  \\
        \hspace{1em}Only Gradle (\texttt{run\_gradle})    & 55.8\%  \\
        \hspace{1em}Only build (\texttt{run\_build})      & 63.4\%  \\
        \midrule
        \multicolumn{2}{@{}l}{\textbf{Tool Combinations}} \\
        \hspace{1em}Build + Gradle  & 69.7\%  \\
        \hspace{1em}Build + Java    & 69.8\%  \\ 
        \hspace{1em}Shell + Build + Gradle + Java & 70.7\% \\
        \hspace{1em}\textbf{Build + Gradle + Java (Ours)} & \textbf{74.0\%} \\
        \bottomrule
    \end{tabular}
    \vspace{-5pt}
\end{table}

\begin{table}[t!]
  \begin{threeparttable}
    \caption{Tool usage frequency (\%) across agent configurations. Agents are: \textbf{Base} (no shell), \textbf{Shell} (base with shell), \textbf{Gradle} (Only gradle), \textbf{Build} (Only Build), and \textbf{Ours} (Gradle, Build, and Java).}
    \label{tab:tool_usage_counts}
    \setlength{\tabcolsep}{3pt} % Adjust column spacing
    \sisetup{table-format=2.1}
    \footnotesize % Use a smaller font
    \begin{tabular*}{\columnwidth}{@{\extracolsep{\fill}} l S S S S S}
        \toprule
        \textbf{Tool} & {\textbf{Base}} & {\textbf{Shell}} & {\textbf{Gradle}} & {\textbf{Build}} & {\textbf{Ours}} \\
        \midrule
        LS                      & 13.0  & 1.0   & 1.9   & 1.1   & 3.2   \\
        Grep                    & 3.2   & 25.7  & 1.3   & 1.3   & 1.4   \\
        Glob                    & 9.6   & 1.3   & 2.2   & 2.5   & 1.8   \\
        Read File               & 35.1  & 19.0  & 32.0  & 32.2  & 33.6  \\
        Replace                 & 39.1  & 5.9   & 36.3  & 39.1  & 41.2  \\
        Google Search           & 0.0   & 0.1   & 1.7   & 3.6   & 1.6   \\
        \midrule
        \textbf{Shell (Total)}  & {-}   & \textbf{47.0}  & {-}   & {-}   & {-}   \\
        \hspace{1em}Change Java & {-} & 13.3\tnote{*} & {-} & {-} & 0.6 \\
        \hspace{1em}Gradle      & {-} & 4.8\tnote{*}  & 24.7 & {-} & 1.8 \\
        \hspace{1em}Build      & {-} & 20.7\tnote{*} & {-} & 20.2 & 14.8 \\
        \bottomrule
    \end{tabular*}
    \begin{tablenotes}[flushleft]
        \item[*] \footnotesize The agent calls these tools using the shell. Percentages for the tools are normalized to represent their fraction of \textbf{all} calls for direct comparability.
    \end{tablenotes}
  \end{threeparttable}
  \vspace{-15pt}
\end{table}

Our results show a clear trend of \textbf{improved performance as tools become more specific}. The agent with only the shell tool improves over the baseline. However, performance increases with the more abstract \texttt{run\_gradle} tool and is highest with the most specific tool, \texttt{run\_build}. This result is particularly notable because \texttt{shell} is a superset of \texttt{run\_gradle}, which is a superset of \texttt{run\_build}. The agent is most effective when given the most constrained, highest-abstraction tool, suggesting the performance gain comes from focus, not just capability. Performance further improved as tools were combined, with our full method (Build+Gradle+Java) achieving the highest resolve rate and demonstrating the value of well-designed shell abstractions. Furthermore, our full method, which excludes the shell, outperforms the combination that includes it (74.0\% vs 70.7\%), suggesting that a constrained and relevant toolset is more effective than an unrestricted one. These observations align with findings in~\citet{wang2023voyager}, where specialized toolsets enable continuous performance improvement while more generic agents plateau. 

These results validate our \textit{Tool Bridging} strategy, which bridges the gap between an LLM's high-level knowledge and its low-level execution. The shell-based agent's behavior suggests such a gap exists. Table \ref{tab:tool_usage_counts} shows it frequently attempts the correct raw commands for building (20.7\%) and changing Java versions (13.3\%), yet its low pass rate (Table \ref{tab:main_results_wide}) demonstrates it struggles to apply them effectively in a raw shell. This pattern suggests the agent possesses the high-level knowledge of \textit{what} operations are needed but struggles to \textit{correctly sequence and apply} those operations. We illustrate these patterns in our case studies where the shell-based agent becomes trapped by misleading logs or reverts correct fixes (Appendix \ref{app:case_studies}).

\textit{Tool Bridging} addresses this gap by providing domain-aware abstractions. By replacing the general shell with specific tools like \texttt{run\_build}, we reframe complex shell interactions as simple, API-like calls. This aligns with the format LLMs are trained to use reliably~\citep{patil2025bfcl} and constrains the action space to relevant operations. This allows the model to leverage its high-level reasoning without failing on low-level command syntax and sequencing. Our results support this hypothesis. As shown in Table \ref{tab:domain_tools}, performance improves as tools become more specific, suggesting LLMs know which commands are necessary but struggles to sequence them correctly as tools become less specific. GradleFixer achieves superior performance while calling \texttt{change\_java\_version} far less frequently (0.6\%), suggesting effective targeted tool use rather than exploratory use.

In contrast to a general-purpose shell, a dedicated tool like \texttt{run\_gradle} could be priming the LLM to Android knowledge not only through its API-like structure but also through its name and description. This explicit contextual signal appears to help the model more reliably apply its latent knowledge of Android builds, steering it away from the incorrect interactions with a generic shell, as demonstrated in our case studies (Appendix \ref{app:case_studies}).

\begin{table}[t!]
\centering
\small
\caption{The Pass@1 resolve rate comparison for Gemini-2.5-Pro and GPT-5-Mini (30 LLM calls per task). }
\label{tab:gemini_performance_booktabs}
\begin{tabular}{lcc}
\toprule
\textbf{Method} & \textbf{Gemini} & \textbf{GPT} \\
\midrule
Shell Agent (Gemini-CLI) & 54.3\%  & 41.5\% \\
+ Tool Usage Guidance   & 58.9\%  & 44.1\% \\
GradleFixer (Ours) & 74.0\% & 59.6\% \\
\bottomrule
\end{tabular}
\vspace{-10pt}
\end{table}

\subsection{Prompting vs. Dedicated Tools}
Given that our domain-specific tools are specialized shell commands, we studied the impact of dedicated tools over prompting methods. We compared the standard shell-enabled Gemini-CLI agent to a version with a "Tool Usage Guidance" prompt that explicitly instructed the LLM on how to perform the build-related tasks using the shell. We added the following instructions to its system prompt:

\begin{quote}
\small
\noindent \textbf{Tool Usage Guidance:}\\
- You can change the Java version by setting the environment variable: \texttt{export JAVA\_HOME=.../[JAVA\_VERSION]/lib/jvm}\\
- You can run specific Gradle tasks using the shell command: \texttt{./gradlew [TASK\_NAME]}\\
- Once you apply a fix, you can test it using the shell command: \texttt{./gradlew assembleDebug}
\end{quote}

We also include an evaluation using GPT-5-Mini to show that our finding is not limited to a single family of LLMs. While OpenAI provides its own coding agent, Codex \cite{openai-codex}, it is not open-source, so its exact toolset cannot be reproduced. We instead use the same tools as Gemini-CLI. The results are shown in Table \ref{tab:gemini_performance_booktabs}. 

We find that providing a tool usage guidance prompt moderately improves performance over the baseline. This shows that guiding the LLM toward important tool use is an effective strategy. However, the prompted agent still performs significantly worse than our proposed agent, GradleFixer, which is equipped with domain-specific tools (Build, Gradle, and Java). 
This performance gap suggests that \textbf{providing domain knowledge through dedicated tools is more effective} than providing it through prompts for a general-purpose tool. We hypothesize this is because dedicated tools present operations in a more reliable, API-like format and constrain the action space, which more effectively connects the agent's high-level reasoning to the low-level execution required for builds.

\begin{figure}[htbp]
    \centering 
    \hspace{-20pt}
    \vspace{-15pt}
    \includegraphics[width=0.9\columnwidth]{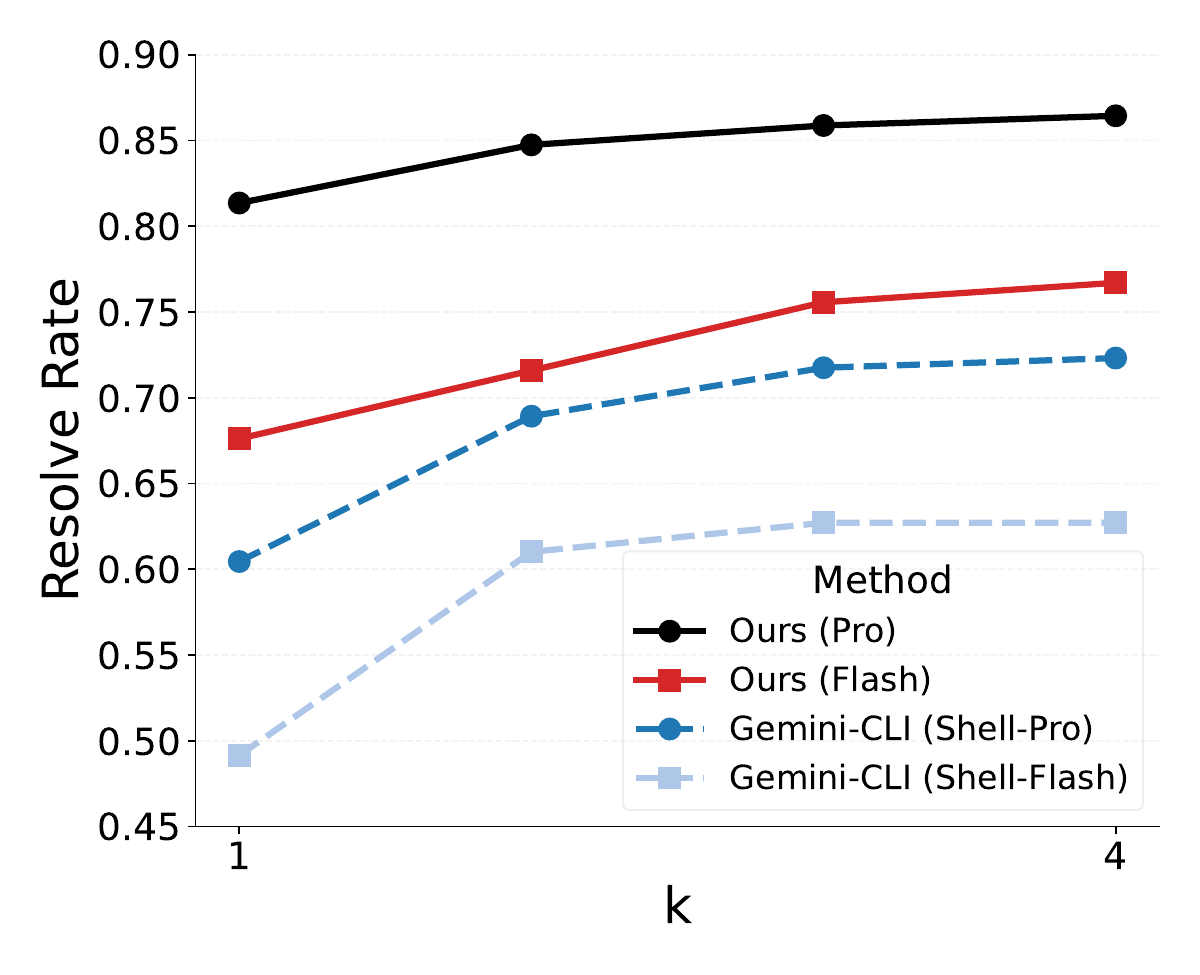}
    \caption{The pass@k resolve rates with a comparison for model sizes. We find that our method, using a smaller model outperforms Gemini-CLI using a larger model, supporting the importance of domain-specific tools on performance. }
    \label{fig:smaller_model}
    \vspace{-10pt}
\end{figure}

% \noindent \textbf{Model Size and Cost-Effectiveness.}
\subsection{Model Size and Cost-Effectiveness}
To study the effect of model size within the same family of LLMs, we evaluated both Gemini-2.5-Pro and the smaller Gemini-2.5-Flash model. We tested both LLMs using our GradleFixer method and the standard Gemini-CLI (Shell) agent. The results are shown in Figure \ref{fig:smaller_model}. We find that GradleFixer with the smaller Flash model outperforms the standard Gemini-CLI agent using the larger Pro model.

This finding suggests that providing useful domain-specific tools can be more impactful than using a larger, more capable model, with significant implications for cost-effectiveness. The Gemini-2.5-Flash model costs less than a quarter of the price per token compared to the Pro model. The cost savings are further increased due to the nature of agent-based repair. When an agent fails, it continues attempting different strategies, consuming many more tokens. In our experiments, successful repairs with GradleFixer averaged 1.1 million input and 10,265 output tokens over 10.6 LLM calls. By contrast, failed attempts consumed nearly four times as many input tokens (4.1 million) and 28,109 output tokens over 27.5 LLM calls. Because a more performant method fails less often, it is also more economical. 

This principle also supports practical cost-saving strategies like a \textit{cascading} agent design, where tasks are first attempted by a smaller, cheaper model and only escalated to a larger one upon failure.
More broadly, this finding points toward a promising future where smaller, specialized models could be fine-tuned to achieve high performance on domain-specific tasks, potentially exceeding the capabilities of larger generalist models at a much lower cost \citep{belcak2025smalllanguagemodelsfuture}.

\begin{table}[t!]
    \centering
    \small
    \caption{Breakdown of success and failure cases. For each metric, we show the performance of Gemini-CLI and our method, GradleFixer. Statistics of the change that led to the failure are reported as Median $\pm$ Std Dev.}
    \label{tab:success_failure_analysis}
    \sisetup{table-format=3.1} 
    \begin{tabular}{l S S}
        \toprule
        \textbf{Metric} & {\textbf{Gemini-CLI}} & {\textbf{GradleFixer}} \\
        \midrule
        \multicolumn{3}{l}{\textit{Root Cause Category (Success Rate)}} \\
        \addlinespace
        \quad Syntax (\%)          & 78.4 & 91.0 \\
        \quad Resource File (\%)   & 92.3 & 100.0 \\
        \quad Configuration (\%)   & 57.8 & 82.2 \\
        \quad Library (\%)         & 88.2 & 88.2 \\
        \addlinespace
        \midrule
        \multicolumn{3}{l}{\textit{Change that led to build error (Fixed Cases)}} \\
        \addlinespace
        \quad Lines Changed        & {80.0 $\pm$ 332.6} & {75.0 $\pm$ 336.3} \\
        \quad Files Changed        & {5.0 $\pm$ 7.8}   & {4.0 $\pm$ 7.4}   \\
        \addlinespace
        \midrule
        \multicolumn{3}{l}{\textit{Change that led to build error (Failed Cases)}} \\
        \addlinespace
        \quad Lines Changed        & {232.0 $\pm$ 677.0} & {714.5 $\pm$ 671.0} \\
        \quad Files Changed        & {8.0 $\pm$ 7.4}    & {15.0 $\pm$ 6.6}   \\
        % \midrule
        % \multicolumn{3}{l}{\textit{Agent's Applied Fix}} \\
        % \addlinespace
        % \quad Lines Changed (Correct)       & {10 $\pm$ 181.8} & {12  $\pm$ 161.7} \\
        % \quad Lines Changed (Incorrect)     & {12 $\pm$ 210.6} & {14 $\pm$ 208.8}   \\
        \bottomrule
    \end{tabular}
    \vspace{-15pt}
\end{table}

\subsection{Analysis of Failure Cases}
While our primary results (Table \ref{tab:main_results_wide}) suggests Dependency-Augmented errors are the most challenging category, a more detailed analysis reveals that the magnitude of the code change leading to the error is a stronger predictor of failure. Table \ref{tab:success_failure_analysis} provides an overview of the performance by error category, and the statistics of code changes for both successful and failed attempts.

The substantially higher median lines and files changed for failed cases in both agents confirms that larger changes are inherently more difficult to fix. This trend is more pronounced for GradleFixer, where the median lines changed in failed cases is nearly ten times higher than successful cases (714.5 vs. 75.0). This shows that while GradleFixer mitigates challenges from different error types, its remaining failures are concentrated in tasks with extensive code modifications, regardless of cause.

This correlation is interesting because the agents do not have access to the original diff that introduced the error. We hypothesize that larger changes are more likely to introduce compounding errors. An initial error can mask subsequent ones, forcing the agent to solve a series of problems iteratively, which significantly increases the task complexity. For example, one case study found that fixing a data binding error revealed a second build configuration issue (Appendix \ref{app:case_study_kapt}).
This finding suggests a clear best practice for developers using automated repair: \textit{building frequently after small, incremental code changes creates an environment where these agents are most likely to succeed, maximizing both effectiveness and cost-efficiency.}

\subsection{Analysis of Tool Semantics}
To study how performance is impacted by semantic priming of tool names (e.g., \texttt{run\_build}) and the functional abstraction of the tool itself, we conducted an ablation study masking tool definitions. We evaluated versions of GradleFixer where tool names were replaced with generic labels (e.g., \texttt{tool\_A}) and descriptions had keywords similarly masked.

As shown in Table \ref{tab:renaming_ablation}, masking tool names resulted in a negligible performance drop, whereas masking descriptions caused a larger drop. However, even with names and descriptions masked, the API-like tool structure achieves a pass@1 rate of 64.9\%, outperforming the shell-based agent and the shell agent with prompt guidance. This indicates that while semantics (particularly descriptions) provide a benefit, the structural constraint of the action space and the reliability of the API-like format also provides significant performance gains.

\begin{table}[h]
\centering
\small
\caption{Ablation study disentangling tool semantics from structure. Even with masked names and descriptions, the specialized tool structure outperforms the generic shell.}
\label{tab:renaming_ablation}
\begin{tabular}{lc}
\toprule
\textbf{Method} & \textbf{Pass@1} \\
\midrule
Shell Agent (Baseline) & 54.3\% \\
Tool Usage Guidance Prompt & 58.9\% \\
\midrule
GradleFixer (Masked Names) & 73.5\% \\
GradleFixer (Masked Descriptions) & 63.4\% \\
GradleFixer (Masked Names \& Desc.) & 64.9\% \\
GradleFixer (Full Semantics) & \textbf{74.0\%} \\
\bottomrule
\end{tabular}
\end{table}

\section{Relation to Prior Work on Domain-specific Tools}
\label{app:relation_to_prior_work}
Our primary contribution is the empirical validation of two agent design insights. First, replacing a general-purpose \texttt{shell} with a curated set of domain-specific tools improves performance. Second, this performance gain increases as the specificity of the tools increases (Table \ref{tab:domain_tools}). We term this strategy \textit{Tool Bridging}.

While we were inspired by prior studies that successfully use domain-specific tools \citep{wang2023voyager, singh2025code}, their focus differs from ours. In prior work, domain-specific tools are a component for achieving a broader goal, such as lifelong learning or deep code analysis. In contrast, our work isolates the impact of domain-specific tools as the central research question and provides insights through empirical evidence.

\noindent\textbf{Distinction from Voyager \citep{wang2023voyager}.}
The Voyager agent is an influential paper for lifelong learning within an embodied environment (Minecraft). It operates by building a growing skill library from executable JavaScript functions which call predefined, domain-specific APIs like \texttt{mineBlock(bot, name)}. These APIs form the agent's action space.

The key difference is that their contribution is not an analysis comparing domain-specific tools against general tools. The novelties are the learned skill library and iterative prompting. The specialized APIs are a prerequisite for the agent to function. However, the paper does not analyze settings where the agent uses a general Minecraft command-line shell, nor does it test if more specific tools would further improve performance. In contrast, we directly compare an agent with a general \texttt{shell} against one with domain-specific tools (GradleFixer), attributing the performance gain to the \textit{Tool Bridging} strategy.

\noindent\textbf{Distinction from Code Researcher \citep{singh2025code}.}
Code Researcher is a powerful agent designed to patch crashes in complex Linux systems code. It uses specialized tools for code analysis, such as \texttt{search\_code(regex)} and \texttt{search\_commits(regex)}, which are domain-specific for code and history investigation.
Similar to Voyager, the paper's contribution is the deep research methodology. The authors do not investigate whether \texttt{search\_commits} is superior to \texttt{git log}, but rather how an agent can use history-aware tools to solve complex bugs. Conversely, our work focuses directly on this comparison. We provide evidence that abstracting complex commands into a constrained, API-like action space is a key strategy for improving agent performance in the Android build domain.

This distinction also explains a key aspect of our experimental design. The goal of our agent is to repair a build error, not simply to identify the commit that introduced it. Because our dataset consists of build-breaking commits, a history-aware tool would allow the agent to solve the problem by reverting the change. We excluded this capability to keep the task focused on code repair.

\section{Conclusion and Future Work}
In conclusion, to address the high frequency of build failures in Android development, we introduced AndroidBuildBench, a benchmark of real-world build errors, and GradleFixer, an LLM agent with domain-specific tools. Our results demonstrate that providing specialized tools is more effective than relying on a general-purpose shell for Android build errors. GradleFixer significantly outperformed state-of-the-art baselines, and we found that a smaller, more cost-effective model with our toolset outperformed a larger model without it. The effectiveness of \textit{Tool Bridging} also offer a potential explanation for the strong performance of agents in other domains that use special API-like tools~\citep{wang2023voyager, singh2025code}. Our work provides the empirical evidence for this design pattern, suggesting it is a generalizable approach for building capable LLM agents.

This work suggests several directions for future research. First, fine-tuning smaller, cost-effective language models on domain-specific datasets like AndroidBuildBench using specialized tools could exceed the performance of larger models. Second, the \textit{Tool Bridging} strategy could be applied to other domains. A compelling research direction is developing agents that automatically generate and refine their own domain-specific tools from experience, allowing them to adapt to specialized tasks without manual tool engineering. Finally, by automating build fixing, our approach could lower the barrier for Android development. This may enable a more fluid and exploratory style of development, such as "vibe-coding," where non-developers and LLMs can experiment with less fear of breaking the build.

% REQUIRED SECTION (DOES NOT COUNT TOWARDS PAGE LIMIT)
\section*{Limitations}
Our study has several limitations. First, AndroidBuildBench is curated from 43 popular, open source Android projects. This selection may not fully represent the diversity of applications, particularly private or less actively maintained projects. Our curation method, which anchors failures to successfully resolved pull requests, also filters out certain persistent environmental issues, such as the NDK errors noted in our analysis, and may therefore underrepresent some categories of build errors.

Second, the GradleFixer agent and its tools are designed for the Android Gradle environment. While we propose \textit{Tool Bridging} as a general strategy, its effectiveness in other development ecosystems, like iOS or web development, has not been tested. Applying this strategy to new domains would require designing a new set of domain-aware tools.

Finally, our explanation for the success of \textit{Tool Bridging} is a hypothesis supported by strong empirical results. Proving the exact internal reasoning of the language model remains a significant challenge in AI research, and our study does not address the deeper, mechanistic question behind our central finding. For instance, inspired by recent work on the internal mechanisms of LLMs \citep{lindsey2025biology}, a deeper investigation could probe whether a dedicated tool like \texttt{run\_build} activates a different, more specialized ``computational circuit'' within the model compared to a generic \texttt{shell} command. Such an analysis would be necessary to connect the behavioral success we observed with its underlying cognitive processes. Therefore, our claim is an inference based on observed performance rather than a direct measurement of the model's cognitive process. Furthermore, our experimental design deliberately excluded access to commit history to create a focused repair task. This differs from a real world scenario where a developer might use version history to diagnose a problem.

\newpage 

\bibliography{custom}

@misc{statista_googleplay_2024,
  author       = {Statista},
  title        = {{Google Play: number of available apps as of Q2 2024}},
  year         = {2024},
  month        = {September},
  howpublished = {\url{https://www.statista.com/statistics/289418/number-of-available-apps-in-the-google-play-store-quarter/}},
  note         = {Accessed: 2025-09-23},
}

@misc{statcounter_os_share,
  author       = {{StatCounter Global Stats}},
  title        = {{Mobile Operating system market share worldwide}},
  howpublished = {\url{https://gs.statcounter.com/os-market-share/mobile/worldwide}},
  note         = {Accessed: 2025-09-23},
}

@misc{stackoverflow_android_questions,
  author       = {{Stack Overflow}},
  title        = {{Newest “Android” questions}},
  howpublished = {\url{https://stackoverflow.com/questions/tagged/android?sort=Newest&edited=true}},
  note         = {Accessed: 2025-09-23},
}

@article{liu2024understanding,
  title={Understanding the quality and evolution of Android app build systems},
  author={Liu, Pei and Li, Li and Liu, Kui and McIntosh, Shane and Grundy, John},
  journal={Journal of Software: Evolution and Process},
  volume={36},
  number={5},
  pages={e2602},
  year={2024},
  publisher={Wiley Online Library}
}

@article{ghaleb2024ci,
  title={CI/CD Configuration Practices in Open-Source Android Apps: An Empirical Study},
  author={Ghaleb, Taher and Abduljalil, Osamah and Hassan, Safwat},
  journal={ACM Transactions on Software Engineering and Methodology},
  year={2024},
  publisher={ACM New York, NY}
}

@inproceedings{hassan2017automatic,
  title={Automatic building of java projects in software repositories: A study on feasibility and challenges},
  author={Hassan, Foyzul and Mostafa, Shaikh and Lam, Edmund SL and Wang, Xiaoyin},
  booktitle={2017 ACM/IEEE International Symposium on Empirical Software Engineering and Measurement (ESEM)},
  pages={38--47},
  year={2017},
  organization={IEEE}
}

@article{jimenez2023swe,
  title={Swe-bench: Can language models resolve real-world github issues?},
  author={Jimenez, Carlos E and Yang, John and Wettig, Alexander and Yao, Shunyu and Pei, Kexin and Press, Ofir and Narasimhan, Karthik},
  journal={arXiv preprint arXiv:2310.06770},
  year={2023}
}

@article{singh2025code,
  title={Code Researcher: Deep Research Agent for Large Systems Code and Commit History},
  author={Singh, Ramneet and Joel, Sathvik and Mehrotra, Abhav and Wadhwa, Nalin and Bairi, Ramakrishna B and Kanade, Aditya and Natarajan, Nagarajan},
  journal={arXiv preprint arXiv:2506.11060},
  year={2025}
}

@article{wang2023voyager,
  title={Voyager: An open-ended embodied agent with large language models},
  author={Wang, Guanzhi and Xie, Yuqi and Jiang, Yunfan and Mandlekar, Ajay and Xiao, Chaowei and Zhu, Yuke and Fan, Linxi and Anandkumar, Anima},
  journal={arXiv preprint arXiv:2305.16291},
  year={2023}
}

@misc{google_gemini_cli,
  author       = {{Google Cloud}},
  title        = {{Gemini CLI}},
  howpublished = {\url{https://cloud.google.com/gemini/docs/codeassist/gemini-cli}},
  note         = {Accessed: 2025-09-23},
}

@article{zhang2025nemotron,
  title={Nemotron-Research-Tool-N1: Exploring Tool-Using Language Models with Reinforced Reasoning},
  author={Zhang, Shaokun and Dong, Yi and Zhang, Jieyu and Kautz, Jan and Catanzaro, Bryan and Tao, Andrew and Wu, Qingyun and Yu, Zhiding and Liu, Guilin},
  journal={arXiv preprint arXiv:2505.00024},
  year={2025}
}

@misc{aider-website,
  author       = {Aider},
  title        = {Aider - AI pair programming in your terminal},
  howpublished = {\url{https://aider.chat/}},
  note         = {(n.d.). Accessed: 2025-09-24},
}

@incollection{kluyver2016jupyter,
  title={Jupyter Notebooks--a publishing format for reproducible computational workflows},
  author={Kluyver, Thomas and Ragan-Kelley, Benjamin and P{\'e}rez, Fernando and Granger, Brian and Bussonnier, Matthias and Frederic, Jonathan and Kelley, Kyle and Hamrick, Jessica and Grout, Jason and Corlay, Sylvain and others},
  booktitle={Positioning and power in academic publishing: Players, agents and agendas},
  pages={87--90},
  year={2016},
  publisher={IOS press}
}

@article{comanici2025gemini,
  title={Gemini 2.5: Pushing the frontier with advanced reasoning, multimodality, long context, and next generation agentic capabilities},
  author={Comanici, Gheorghe and Bieber, Eric and Schaekermann, Mike and Pasupat, Ice and Sachdeva, Noveen and Dhillon, Inderjit and Blistein, Marcel and Ram, Ori and Zhang, Dan and Rosen, Evan and others},
  journal={arXiv preprint arXiv:2507.06261},
  year={2025}
}

@misc{openai-codex,
  author       = {{OpenAI}},
  title        = {Introducing Codex},
  year         = {2025},
  howpublished = {\url{https://openai.com/index/introducing-codex/}},
  note         = {Accessed: 2025-09-25},
}

@misc{belcak2025smalllanguagemodelsfuture,
      title={Small Language Models are the Future of Agentic AI}, 
      author={Peter Belcak and Greg Heinrich and Shizhe Diao and Yonggan Fu and Xin Dong and Saurav Muralidharan and Yingyan Celine Lin and Pavlo Molchanov},
      year={2025},
      eprint={2506.02153},
      archivePrefix={arXiv},
      primaryClass={cs.AI},
      url={https://arxiv.org/abs/2506.02153}, 
}

@article{rostami2023usage,
  title={On the usage, co-usage and migration of CI/CD tools: A qualitative analysis},
  author={Rostami Mazrae, Pooya and Mens, Tom and Golzadeh, Mehdi and Decan, Alexandre},
  journal={Empirical Software Engineering},
  volume={28},
  number={2},
  pages={52},
  year={2023},
  publisher={Springer}
}

@inproceedings{baitha2024streamlining,
  title={Streamlining Software Development: a comprehensive study on CI/CD automation},
  author={Baitha, Sanjay and Soorya, Vijayakumar and Kothari, Osho and Rajagopal, Shinu M and Panda, Niharika},
  booktitle={2024 4th International Conference on Sustainable Expert Systems (ICSES)},
  pages={1299--1305},
  year={2024},
  organization={IEEE}
}

@inproceedings{liu2022first,
  title={A first look at CI/CD adoptions in open-source android apps},
  author={Liu, Pei and Sun, Xiaoyu and Zhao, Yanjie and Liu, Yonghui and Grundy, John and Li, Li},
  booktitle={Proceedings of the 37th IEEE/ACM International Conference on Automated Software Engineering},
  pages={1--6},
  year={2022}
}

@article{zhang2024systematic,
  title={A systematic literature review on large language models for automated program repair},
  author={Zhang, Quanjun and Fang, Chunrong and Xie, Yang and Ma, YuXiang and Sun, Weisong and Yang, Yun and Chen, Zhenyu},
  journal={arXiv preprint arXiv:2405.01466},
  year={2024}
}

@inproceedings{yin2024thinkrepair,
  title={Thinkrepair: Self-directed automated program repair},
  author={Yin, Xin and Ni, Chao and Wang, Shaohua and Li, Zhenhao and Zeng, Limin and Yang, Xiaohu},
  booktitle={Proceedings of the 33rd ACM SIGSOFT International Symposium on Software Testing and Analysis},
  pages={1274--1286},
  year={2024}
}

@inproceedings{xia2023automated,
  title={Automated program repair in the era of large pre-trained language models},
  author={Xia, Chunqiu Steven and Wei, Yuxiang and Zhang, Lingming},
  booktitle={2023 IEEE/ACM 45th International Conference on Software Engineering (ICSE)},
  pages={1482--1494},
  year={2023},
  organization={IEEE}
}

@inproceedings{jin2023inferfix,
  title={Inferfix: End-to-end program repair with llms},
  author={Jin, Matthew and Shahriar, Syed and Tufano, Michele and Shi, Xin and Lu, Shuai and Sundaresan, Neel and Svyatkovskiy, Alexey},
  booktitle={Proceedings of the 31st ACM joint european software engineering conference and symposium on the foundations of software engineering},
  pages={1646--1656},
  year={2023}
}

@article{tambon2025bugs,
  title={Bugs in large language models generated code: An empirical study},
  author={Tambon, Florian and Moradi-Dakhel, Arghavan and Nikanjam, Amin and Khomh, Foutse and Desmarais, Michel C and Antoniol, Giuliano},
  journal={Empirical Software Engineering},
  volume={30},
  number={3},
  pages={65},
  year={2025},
  publisher={Springer}
}

@inproceedings{patil2025bfcl,
title={The Berkeley Function Calling Leaderboard (BFCL): From Tool Use to Agentic Evaluation of Large Language Models}, 
author={Patil, Shishir G. and Mao, Huanzhi and Cheng-Jie Ji, Charlie and Yan, Fanjia and Suresh, Vishnu and Stoica, Ion and E. Gonzalez, Joseph},
booktitle={Forty-second International Conference on Machine Learning},
year={2025},
}

@article{liu2024toolace,
  title={Toolace: Winning the points of llm function calling},
  author={Liu, Weiwen and Huang, Xu and Zeng, Xingshan and Hao, Xinlong and Yu, Shuai and Li, Dexun and Wang, Shuai and Gan, Weinan and Liu, Zhengying and Yu, Yuanqing and others},
  journal={arXiv preprint arXiv:2409.00920},
  year={2024}
}

@article{zhang2024ecoact,
  title={Ecoact: Economic agent determines when to register what action},
  author={Zhang, Shaokun and Zhang, Jieyu and Ding, Dujian and Garcia, Mirian Hipolito and Mallick, Ankur and Madrigal, Daniel and Xia, Menglin and R{\"u}hle, Victor and Wu, Qingyun and Wang, Chi},
  journal={arXiv preprint arXiv:2411.01643},
  year={2024}
}

@article{patil2024gorilla,
  title={Gorilla: Large language model connected with massive apis},
  author={Patil, Shishir G and Zhang, Tianjun and Wang, Xin and Gonzalez, Joseph E},
  journal={Advances in Neural Information Processing Systems},
  volume={37},
  pages={126544--126565},
  year={2024}
}

@article{ishibashi2024self,
  title={Self-organized agents: A llm multi-agent framework toward ultra large-scale code generation and optimization},
  author={Ishibashi, Yoichi and Nishimura, Yoshimasa},
  journal={arXiv preprint arXiv:2404.02183},
  year={2024}
}

@article{liu2024towards,
  title={Towards hierarchical multi-agent workflows for zero-shot prompt optimization},
  author={Liu, Yuchi and Singh, Jaskirat and Liu, Gaowen and Payani, Ali and Zheng, Liang},
  journal={arXiv preprint arXiv:2405.20252},
  year={2024}
}

@article{zeng2025glm,
  title={Glm-4.5: Agentic, reasoning, and coding (arc) foundation models},
  author={Zeng, Aohan and Lv, Xin and Zheng, Qinkai and Hou, Zhenyu and Chen, Bin and Xie, Chengxing and Wang, Cunxiang and Yin, Da and Zeng, Hao and Zhang, Jiajie and others},
  journal={arXiv preprint arXiv:2508.06471},
  year={2025}
}

@misc{tbench_2025,
      title={Terminal-Bench: A Benchmark for AI Agents in Terminal Environments}, 
      url={https://github.com/laude-institute/terminal-bench}, 
      author={The Terminal-Bench Team}, year={2025}, month={Apr}}

@article{lindsey2025biology,
  author={Lindsey, Jack and Gurnee, Wes and Ameisen, Emmanuel and Chen, Brian and Pearce, Adam and Turner, Nicholas L. and Citro, Craig and Abrahams, David and Carter, Shan and Hosmer, Basil and Marcus, Jonathan and Sklar, Michael and Templeton, Adly and Bricken, Trenton and McDougall, Callum and Cunningham, Hoagy and Henighan, Thomas and Jermyn, Adam and Jones, Andy and Persic, Andrew and Qi, Zhenyi and Thompson, T. Ben and Zimmerman, Sam and Rivoire, Kelley and Conerly, Thomas and Olah, Chris and Batson, Joshua},
  title={On the Biology of a Large Language Model},
  journal={Transformer Circuits Thread},
  year={2025},
  url={https://transformer-circuits.pub/2025/attribution-graphs/biology.html}
}

\appendix

\begin{table}[h!]
\centering
\caption{Comparison of the official Gemini-CLI agent and our Python replica. Both agents were configured with only the \texttt{shell} tool. The similar pass@1 rates validate the accuracy of our replica.}
\label{tab:replica_validation}
\begin{tabular}{lc}
\toprule
\textbf{Implementation} & \textbf{Pass@1 (\%)} \\
\midrule
Original Gemini-CLI & 65.1 \\
Our Python Replica & 65.7 \\
\bottomrule
\end{tabular}
\end{table}

\section{Validation of the Gemini-CLI Replica}
\label{app:replica_validation}
In our experiments, we required the ability to systematically modify the toolset available to the agent to isolate the impact of domain-specific tools. To facilitate these controlled experiments, we developed a Python replica of the open-source Gemini-CLI agent \citep{google_gemini_cli}. Our replica preserves the core agentic loop and prompting structure of the original implementation while allowing for straightforward customization of the available tools.
To validate the fidelity of our replica, we conducted an experiment to compare its performance against the official, unmodified Gemini-CLI agent. For a fair comparison, both agents were configured to use only the general-purpose \texttt{shell} tool, in addition to standard file system tools. This configuration tests whether our replica preserves the capabilities of the original agent framework.
The results of this validation are shown in Table \ref{tab:replica_validation}. We find that the resolve rates are nearly identical. This small difference is well within the expected variance due to model stochasticity. Temperature is kept at the default of 1. The close alignment in performance confirms that our replica is an accurate and reliable reproduction of the original agent, validating its use for the \textbf{Gemini-CLI (Read/Write Only)} and \textbf{GradleFixer} configurations in our experiments.

\section{Prompts}
\label{appendix:prompts}

\subsection{System prompt}
\label{appendix:system_prompt}
Due to the long length of the system prompt, we point readers to the open-source Gemini-CLI repository on GitHub (\texttt{packages/core/src/core/prompts.ts}).

\subsection{Initial prompt}
\label{appendix:initial_prompt}
The following is the initial prompt that is given to the LLM agents.

\begin{lstlisting}[language=bash, caption={Main Prompt}, label={lst:prompt}]

** Current project full path. **
===============================
{os.getcwd()}

**Directory tree:**
===============================
{tree}

** Current State (Build Error):**
===============================
{cur_builderror}
\end{lstlisting}

\newpage

\section{Tools}
\label{appendix:tools}

We provide the tool name, description, and the parameters, which are all provided to the LLM.
\subsection*{Our Tools}
\begin{list}{}{%
\setlength{\leftmargin}{0pt}%
\setlength{\itemsep}{6pt}%
\setlength{\parsep}{0pt}%
}
\item \textbf{1. \texttt{change\_java\_version}}
Changes the active Java version in the environment. This is crucial for projects that require a specific JDK version to build.
\begin{itemize}[nosep, leftmargin=1.5em, topsep=2pt]
\item \textbf{Parameters}:
\item \texttt{version} (integer, required): The target Java version. Must be one of the available versions: [23, 22, 21, 20, 17, 11].
\end{itemize}
\item \textbf{2. \texttt{run\_gradle\_command}}
Executes a specified Gradle command to gather more information about the build process, such as dependency trees or detailed error logs.
\begin{itemize}[nosep, leftmargin=1.5em, topsep=2pt]
\item \textbf{Parameters}:
\item \texttt{gradle\_command} (string, required): The Gradle command to run, e.g., \texttt{./gradlew dependencies}.
\end{itemize}
\item \textbf{3. \texttt{run\_build}}
Initiates a build process using an external validator to test changes and identify any resulting compilation or build errors. Takes no parameters.
\end{list}
\subsection*{Gemini-CLI Original Tools}
\begin{list}{}{%
\setlength{\leftmargin}{0pt}%
\setlength{\itemsep}{6pt}%
\setlength{\parsep}{0pt}%
}
\item \textbf{1. \texttt{run\_shell}}
Executes an arbitrary shell command, providing a flexible interface for interacting with the file system and operating environment.
\begin{itemize}[nosep, leftmargin=1.5em, topsep=2pt]
\item \textbf{Parameters}:
\item \texttt{shell\_command} (string, required): The shell command to be executed.
\end{itemize}
\item \textbf{2. \texttt{replace}}
Performs a find-and-replace operation on a file. It requires precise contextual information and should be preceded by using \texttt{read\_file} to inspect the content.
\begin{itemize}[nosep, leftmargin=1.5em, topsep=2pt]
\item \textbf{Parameters}:
\item \texttt{file\_path} (string, required): Path to the file to be modified.
\item \texttt{old\_string} (string, required): The exact, literal text to be replaced, including surrounding context.
\item \texttt{new\_string} (string, required): The exact, literal text to substitute for \texttt{old\_string}.
\item \texttt{expected\_replacements} (integer, optional): Specifies the number of occurrences to replace (defaults to 1).
\end{itemize}
\item \textbf{3. \texttt{search\_file\_content}}
Searches for a regular expression pattern within files, returning matching lines with their file paths and line numbers.
\begin{itemize}[nosep, leftmargin=1.5em, topsep=2pt]
\item \textbf{Parameters}:
\item \texttt{pattern} (string, required): The regular expression to search for.
\item \texttt{path} (string, optional): The directory to search within.
\item \texttt{include} (string, optional): A glob pattern to filter files (e.g., \texttt{'*.java'}).
\end{itemize}
\item \textbf{4. \texttt{glob}}
Finds files matching a glob pattern, sorted by modification time. Ideal for locating recently changed files.
\begin{itemize}[nosep, leftmargin=1.5em, topsep=2pt]
\item \textbf{Parameters}:
\item \texttt{pattern} (string, required): The glob pattern to match.
\item \texttt{path} (string, optional): The directory to search within.
\item \texttt{case\_sensitive} (boolean, optional): Sets if the match is case-sensitive.
\end{itemize}
\item \textbf{5. \texttt{read\_file}}
Reads the content of a file (text, images, PDFs). For large files, content can be read in chunks using an offset and limit.
\begin{itemize}[nosep, leftmargin=1.5em, topsep=2pt]
\item \textbf{Parameters}:
\item \texttt{path} (string, required): Path of the file to read.
\item \texttt{offset} (integer, optional): The line number to start reading from.
\item \texttt{limit} (integer, optional): The maximum number of lines to read.
\end{itemize}
\item \textbf{6. \texttt{list\_directory}}
Lists the contents (files and subdirectories) of a specified directory.
\begin{itemize}[nosep, leftmargin=1.5em, topsep=2pt]
\item \textbf{Parameters}:
\item \texttt{path} (string, required): The path of the directory to list.
\item \texttt{ignore} (array, optional): A list of glob patterns to exclude from results.
\end{itemize}
\item \textbf{7. \texttt{search\_google}}
Performs a Google search to find solutions for build errors or other issues.
\begin{itemize}[nosep, leftmargin=1.5em, topsep=2pt]
\item \textbf{Parameters}:
\item \texttt{query} (string, required): The search query, typically an error message.
\end{itemize}
\end{list}

\newpage

\begin{table}[htbp]
  \centering
  \small
  \caption{Distribution of problem difficulty, categorized by the number of lines changed in the ground-truth fix.}
  \label{tab:difficulty_distribution}
  \begin{tabular}{lrr}
    \toprule
    \textbf{Difficulty Tier} & \textbf{Count} & \textbf{Percentage (\%)} \\
    \midrule
    Trivial (1--10 lines)   & 351 & 34.5 \\
    Small (11--100 lines)   & 386 & 37.9 \\
    Medium (101--1000 lines) & 226 & 22.2 \\
    Large (>1000 lines)     & 55  & 5.4  \\
    \bottomrule
  \end{tabular}
\end{table}
\section{Distribution of Problem Difficulty}
\label{appendix:problem_diff}

To further characterize the complexity of the problems within AndroidBuildBench, we analyzed the distribution of the number of lines changed that led to the build error, which serves as a proxy for task difficulty. The distribution is highly right-skewed, with the median number of lines changed being 28, and the mean is over ten times higher at 325.2. This disparity shows that the benchmark contains many difficult cases. The most challenging fixes are exceptionally large, with the top 5\% of problems that had 1,000 lines changed.

\begin{figure}[htbp]
    \centering
    \includegraphics[width=0.9\columnwidth]{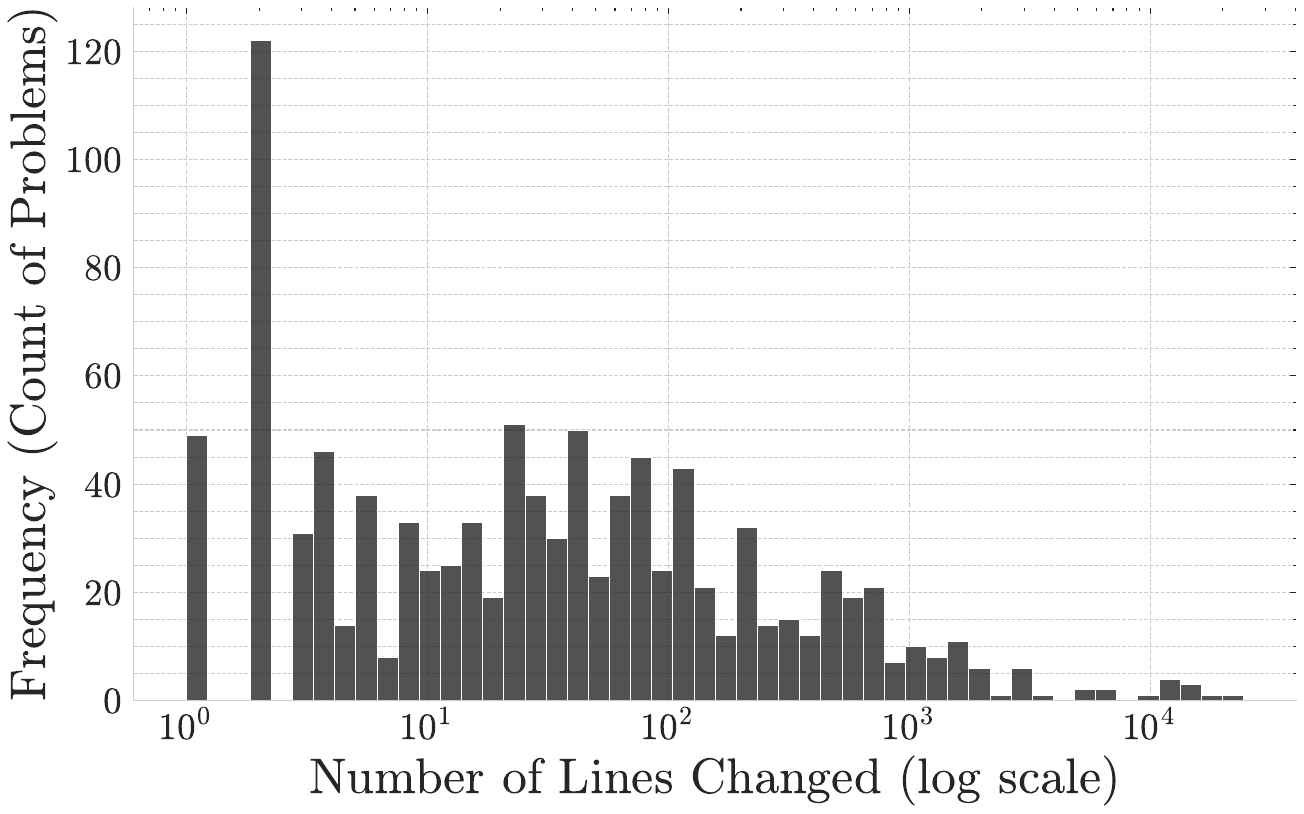} % Replace with the actual path to your histogram image
    \caption{Distribution of problems by the number of lines changed, plotted on a logarithmic x-axis. The histogram shows that while most problems involve small changes, the benchmark contains a long tail of complex problems that had changes to hundreds or thousands of lines.}
    \label{fig:difficulty_histogram}
\end{figure}

Figure \ref{fig:difficulty_histogram} visualizes this distribution on a logarithmic scale. This figure shows that while a majority of fixes are small, the benchmark contains a significant number of problems that have had hundreds or even thousands of lines of code changed. For a clearer summary, Table \ref{tab:difficulty_distribution} categorizes each problem into one of four difficulty tiers. The breakdown shows that a substantial portion of the benchmark consists of `Medium' (22.2\%) and `Large' (5.4\%) problems. This balanced distribution ensures that AndroidBuildBench provides a robust benchmark for evaluating an agent's ability to handle both common, small fixes and the more complex, large-scale modifications representative of challenging real-world software repair tasks.

\section{Case Study Trajectories}
\label{app:case_studies}

This appendix provides detailed step-by-step trajectories for the agents on select tasks, illustrating the differences in their problem-solving approaches. Case studies were randomly selected within curated set of problems that were solved by Gradle Fixer, but were unsolved by Gemini-CLI.

\noindent \textbf{Note on Presentation:} The trajectories presented below are not verbatim outputs from the language models. Instead, they are concise summaries reconstructed from the agents' detailed execution logs. \textit{Step} sections represents our interpretation of what the LLMs wanted to accomplish based on their thoughts and outputs. \textit{Thought} sections represent the agent's reasoning and planning at each stage, while \textit{Tool Call} and \textit{Result} summarize the specific actions taken and their outcomes. This format is designed to highlight the key decision-making points in each agent's process.

\subsection{Case 1: Propagating a Missing Parameter}
\label{app:case_study_propagation}

This case study examines a common build failure caused by a missing function parameter. Resolving this error requires the agent to trace the dependency through multiple files and apply fixes along the entire call chain. It highlights the difference between GradleFixer's systematic, top-down approach and the shell-based agent's more localized but ultimately incorrect strategy.

\subsection*{Build Error Log}
Both agents were presented with the following build failure. The error log clearly indicates that a composable called within \texttt{CommentReplyScreen.kt} is missing a required \texttt{focusManager} parameter.

\begin{lstlisting}[language=bash, caption={Initial Build Failure}, label={lst:build_error_2}, basicstyle=\ttfamily\scriptsize]
> Task :app:compileDebugKotlin FAILED
e: file:///.../app/src/main/java/com/jerboa/ui/components/comment/reply/CommentReplyScreen.kt:73:29: No value passed for parameter 'focusManager'

FAILURE: Build failed with an exception.

* What went wrong:
Execution failed for task ':app:compileDebugKotlin'.
> A failure occurred while executing org.jetbrains.kotlin.compilerRunner.GradleCompilerRunnerWithWorkers$
        GradleKotlinCompilerWorkAction
   > Compilation error. See log for more details
\end{lstlisting}

\subsubsection{GradleFixer (Successful Trajectory)}
GradleFixer correctly identifies that the error requires a multi-step fix that propagates through the codebase. It formulates a high-level plan to modify the entire call chain, from the top-level screen down to the lowest-level component.

\begin{successbox}[Successful Trajectory: GradleFixer]
\begin{enumerate}[label=\textbf{Step \arabic*:}, wide, labelwidth=!, labelindent=0pt, topsep=8pt, itemsep=10pt]

\item \textbf{Diagnose and Apply Initial Fix}
    \begin{itemize}[leftmargin=4.0em, labelindent=0pt, itemsep=3pt, topsep=3pt, parsep=2pt]
        \item[\textbf{Thought:}] The error `No value passed for parameter 'focusManager'` is in \texttt{CommentReplyScreen.kt}. I need to obtain an instance of \texttt{FocusManager} using \texttt{LocalFocusManager.current} and pass it to the composables being called inside the \texttt{when} block.
        \item[\textbf{Tool Call:}] \lstinline|read_file(path='.../CommentReplyScreen.kt')|
        \item[\textbf{Tool Call:}] \lstinline|replace(..., new_string='val focusManager = LocalFocusManager.current')|
        \item[\textbf{Tool Call:}] \lstinline|replace(..., new_string='CommentReply(..., focusManager = focusManager, ...)')|
    \end{itemize}

\item \textbf{Propagate the Fix to Function Definitions}
    \begin{itemize}[leftmargin=4.0em, labelindent=0pt, itemsep=3pt, topsep=3pt, parsep=2pt]
        \item[\textbf{Thought:}] Passing a new parameter will cause a new build error at the function definition site. I must now find the definitions for \texttt{CommentReply}, \texttt{PostReply}, etc., and add the \texttt{focusManager} parameter to their signatures.
        \item[\textbf{Tool Call:}] \lstinline|search_file_content(pattern='fun CommentReply(')|
        \item[\textbf{Result:}] Found definitions in \texttt{.../CommentReply.kt}.
        \item[\textbf{Tool Call:}] \lstinline|replace(file_path='.../CommentReply.kt', ..., new_string='fun CommentReply(..., focusManager: FocusManager, ...)')|
    \end{itemize}

\item \textbf{Trace the Dependency Chain Downward}
    \begin{itemize}[leftmargin=4.0em, labelindent=0pt, itemsep=3pt, topsep=3pt, parsep=2pt]
        \item[\textbf{Thought:}] The \texttt{...Reply} composables use a shared \texttt{MarkdownTextField} composable. To complete the fix, I must also pass the \texttt{focusManager} down to it.
        \item[\textbf{Tool Call:}] \lstinline|replace(file_path='.../CommentReply.kt', ..., new_string='MarkdownTextField(..., focusManager = focusManager, ...)')|
    \end{itemize}

\item \textbf{Modify the Lowest-Level Component}
    \begin{itemize}[leftmargin=4.0em, labelindent=0pt, itemsep=3pt, topsep=3pt, parsep=2pt]
        \item[\textbf{Thought:}] Finally, I need to modify the definition of \texttt{MarkdownTextField}. I'll find its file and add the parameter. I will make it an optional parameter to avoid breaking other usages of this common component throughout the app.
        \item[\textbf{Tool Call:}] \lstinline|search_file_content(pattern='fun MarkdownTextField(')|
        \item[\textbf{Result:}] Found definition in \texttt{.../InputFields.kt}.
        \item[\textbf{Tool Call:}] \lstinline|replace(file_path='.../InputFields.kt', ..., new_string='fun MarkdownTextField(..., focusManager: FocusManager? = null, ...)')|
    \end{itemize}
    
\item \textbf{Verify the Final Fix}
    \begin{itemize}[leftmargin=4.0em, labelindent=0pt, itemsep=3pt, topsep=3pt, parsep=2pt]
        \item[\textbf{Thought:}] I have now updated the entire dependency chain from the call site in \texttt{CommentReplyScreen} down to the final use in \texttt{MarkdownTextField}. I will run the build to confirm the fix.
        \item[\textbf{Tool Call:}] \lstinline|run_build()|
        \item[\textbf{Result:}] {\ttfamily BUILD SUCCESSFUL}. The agent successfully traced and fixed a multi-file dependency issue.
    \end{itemize}
    
\end{enumerate}
\end{successbox}

\subsubsection{Gemini-CLI with Shell (Failed Trajectory)}
The shell-based agent also identifies the error's location but makes a critical reasoning error. It assumes the existing function signatures are correct and immutable, preventing it from considering the necessary top-down fix. This leads it to attempt incorrect, localized patches that ultimately fail.

\begin{failurebox}[Failed Trajectory: Gemini-CLI with Shell]
\begin{enumerate}[label=\textbf{Step \arabic*:}, wide, labelwidth=!, labelindent=0pt, topsep=8pt, itemsep=10pt]

\item \textbf{Initial Diagnosis}
    \begin{itemize}[leftmargin=4.0em, labelindent=0pt, itemsep=3pt, topsep=3pt, parsep=2pt]
        \item[\textbf{Thought:}] The error is in \texttt{CommentReplyScreen.kt}. I need to see the definitions of the functions it is calling to understand why the build is failing.
        \item[\textbf{Tool Call:}] \lstinline|run_shell(shell_command='grep -r "fun PostReply" ...')|
        \item[\textbf{Tool Call:}] \lstinline|read_file(path='.../CommentReply.kt')|
    \end{itemize}

\item \textbf{Make an Incorrect Assumption}
    \begin{itemize}[leftmargin=4.0em, labelindent=0pt, itemsep=3pt, topsep=3pt, parsep=2pt]
        \item[\textbf{Thought:}] I have read the contents of \texttt{CommentReply.kt}. The functions \texttt{CommentReply}, \texttt{PostReply}, etc., do not have a \texttt{focusManager} parameter. The error must be a mistake in the compiler output, or I am misinterpreting it. I should not change the function signatures. The problem is likely in a component used by these functions.
    \end{itemize}

\item \textbf{Focus on a Lower-Level Component}
    \begin{itemize}[leftmargin=4.0em, labelindent=0pt, itemsep=3pt, topsep=3pt, parsep=2pt]
        \item[\textbf{Thought:}] The \texttt{...Reply} composables all use \texttt{MarkdownTextField}. Maybe that is where the problem is. I'll add the \texttt{focusManager} directly to \texttt{MarkdownTextField} and see if that fixes it.
        \item[\textbf{Tool Call:}] \lstinline|run_shell(shell_command='grep -r "fun MarkdownTextField" ...')|
        \item[\textbf{Tool Call:}] \lstinline|replace(file_path='.../InputFields.kt', ..., new_string='fun MarkdownTextField(..., focusManager: FocusManager, ...)')|
    \end{itemize}
    
\item \textbf{Attempt an Invalid Local Patch}
    \begin{itemize}[leftmargin=4.0em, labelindent=0pt, itemsep=3pt, topsep=3pt, parsep=2pt]
        \item[\textbf{Thought:}] Now that \texttt{MarkdownTextField} has the parameter, I will try to pass a \texttt{focusManager} to it from inside \texttt{CommentReply.kt}, but without changing the signature of the \texttt{CommentReply} function itself. I'll just create a new \texttt{focusManager} locally within the function body.
        \item[\textbf{Tool Call:}] \lstinline|replace(file_path='.../CommentReply.kt', ..., new_string='val focusManager = LocalFocusManager.current; MarkdownTextField(..., focusManager = focusManager, ...)')|
    \end{itemize}
    
\item \textbf{Encounter Execution Failures and Give Up}
    \begin{itemize}[leftmargin=4.0em, labelindent=0pt, itemsep=3pt, topsep=3pt, parsep=2pt]
        \item[\textbf{Thought:}] My `replace` calls are failing because the file content does not match what I expected. The build still fails with the same `No value passed` error. My strategy of fixing the lower-level component is not working. I am stuck.
        \item[\textbf{Result:}] The agent gets caught in a loop of reading files and attempting incorrect replacements. It fails to recognize the need to modify function signatures in the middle of the call chain and terminates after reaching its attempt limit, leaving the build broken.
    \end{itemize}

\end{enumerate}
\end{failurebox}

\subsection{Case 2: Misleading Kapt Error and Layered Fixes}
\label{app:case_study_kapt}

This case study illustrates the key finding that vague error log combined with a misleading warning can send a general-purpose agent into an unrecoverable state. In contrast, an agent with domain-specific tools better diagnoses the true root cause and methodically resolve the issue, even when it requires solving layered errors.

\subsection*{Build Error Log}
Both agents were presented with the following build failure. The error is a generic \texttt{kaptDebugKotlin} failure, but the log also contains a prominent warning about Kotlin language version 2.0, which acts as a red herring, suggesting a systemic versioning problem.

\begin{lstlisting}[language=bash, caption={Initial Build Failure}, label={lst:build_error}, basicstyle=\ttfamily\scriptsize]
> Task :app:kaptGenerateStubsDebugKotlin
w: Kapt currently doesn't support language version 2.0+. Falling back to 1.9.

> Task :app:kaptDebugKotlin FAILED

FAILURE: Build failed with an exception.

* What went wrong:
Execution failed for task ':app:kaptDebugKotlin'.
> A failure occurred while executing org.jetbrains.kotlin.gradle.internal.KaptWithoutKotlincTask$KaptExecutionWorkAction

* Try:
> Run with --stacktrace option to get the stack trace.
\end{lstlisting}

\subsubsection{GradleFixer (Successful Trajectory)}
GradleFixer demonstrates a robust, multi-step debugging process. Although it initially explores the misleading versioning issue, it quickly pivots to a more systematic approach, using detailed logs to uncover a chain of two distinct problems and solving them sequentially.

\begin{successbox}[Successful Trajectory: GradleFixer]
\begin{enumerate}[label=\textbf{Step \arabic*:}, wide, labelwidth=!, labelindent=0pt, topsep=8pt, itemsep=10pt]

\item \textbf{Get Detailed Error Information}
    \begin{itemize}[leftmargin=4.0em, labelindent=0pt, itemsep=3pt, topsep=3pt, parsep=2pt]
        \item[\textbf{Thought:}] The initial error is too generic. The warning about Kotlin 2.0 is a possibility, but before I start changing versions, I should follow the log's advice and get a proper stack trace. This will give me the precise origin of the Kapt failure.
        \item[\textbf{Tool Call:}] \lstinline|run_gradle_command(gradle_command='./gradlew assembleDebug --stacktrace')|
    \end{itemize}

\item \textbf{Identify the True Root Cause: Data Binding}
    \begin{itemize}[leftmargin=4.0em, labelindent=0pt, itemsep=3pt, topsep=3pt, parsep=2pt]
        \item[\textbf{Thought:}] The stack trace is perfect. It bypasses the misleading version warning and points to the real issue: a {\ttfamily DataBinding Error}. It says {\ttfamily Cannot find a setter for <android.widget.TextView android:text> that accepts parameter type 'error.NonExistentClass'}. The error is in \url{app/src/main/res/layout/fragment_about.xml}. I will inspect this file.
        \item[\textbf{Tool Call:}] \lstinline|read_file(path='app/src/main/res/layout/fragment_about.xml')|
    \end{itemize}

\item \textbf{Apply a Fix and Uncover a Second Error}
    \begin{itemize}[leftmargin=4.0em, labelindent=0pt, itemsep=3pt, topsep=3pt, parsep=2pt]
        \item[\textbf{Thought:}] I see the problem: \url{android:text="@{context.version, default=0}"}. The `version` is a String, but the `default` is an Integer, causing a type mismatch. I'll fix this by using `String.valueOf()` to ensure type safety.
        \item[\textbf{Tool Call:}] \lstinline|replace(..., new_string='android:text="@{String.valueOf(context.version)}" />')|
        \item[\textbf{Result:}] The build fails again, but with a *new* error: {\ttfamily Unresolved reference: BuildConfig}. This is progress! My data binding fix was correct, but it revealed a second, underlying problem.
    \end{itemize}
    
\item \textbf{Solve the Second Error}
    \begin{itemize}[leftmargin=4.0em, labelindent=0pt, itemsep=3pt, topsep=3pt, parsep=2pt]
        \item[\textbf{Thought:}] The `BuildConfig` class is not being generated. This often happens when build features are misconfigured. I remember an earlier, incorrect attempt where I forced Kotlin's `languageVersion` to 1.9. That must be the cause. I'll revert that change and explicitly enable `buildConfig` generation in the `build.gradle.kts` file, which is the proper solution.
        \item[\textbf{Tool Call:}] \lstinline|replace(..., new_string='buildFeatures { buildConfig = true }', old_string='kotlinOptions { languageVersion = "1.9" }')|
    \end{itemize}
    
\item \textbf{Verify the Final Fix}
    \begin{itemize}[leftmargin=4.0em, labelindent=0pt, itemsep=3pt, topsep=3pt, parsep=2pt]
        \item[\textbf{Thought:}] I have now addressed both the data binding error and the `BuildConfig` generation issue. The final step is to run the build to confirm that all issues are resolved.
        \item[\textbf{Tool Call:}] \lstinline|run_build()|
        \item[\textbf{Result:}] {\ttfamily BUILD SUCCESSFUL}. The agent solved a complex, two-layer problem and terminates.
    \end{itemize}
    
\end{enumerate}
\end{successbox}

\subsubsection{Gemini-CLI with Shell (Failed Trajectory)}
The shell-based agent also successfully identifies the initial root cause. However, when its first fix fails to resolve the issue, it incorrectly abandons this line of investigation. It latches onto the misleading version warning from the initial log and descends into a chaotic and unproductive "version hell," from which it never recovers.

\begin{failurebox}[Failed Trajectory: Gemini-CLI with Shell]
\begin{enumerate}[label=\textbf{Step \arabic*:}, wide, labelwidth=!, labelindent=0pt, topsep=8pt, itemsep=10pt]

\item \textbf{Get Detailed Error Information}
    \begin{itemize}[leftmargin=4.0em, labelindent=0pt, itemsep=3pt, topsep=3pt, parsep=2pt]
        \item[\textbf{Thought:}] The build failed with a Kapt error. I need more information. I'll use the shell to run the build with the \url{--stacktrace} flag as suggested.
        \item[\textbf{Tool Call:}] \lstinline|run_shell(shell_command='./gradlew assembleDebug --stacktrace')|
    \end{itemize}

\item \textbf{Identify the Root Cause and Apply a Plausible Fix}
    \begin{itemize}[leftmargin=4.0em, labelindent=0pt, itemsep=3pt, topsep=3pt, parsep=2pt]
        \item[\textbf{Thought:}] The stack trace clearly shows a data binding error in \url{fragment_about.xml} caused by a type mismatch on a `TextView`. The line has a `default=0` integer value for a string attribute. This is the obvious error. I will remove the default value.
        \item[\textbf{Tool Call:}] \lstinline|replace(..., old_string='android:text="@{context.version, default=0}"', new_string='android:text="@{context.version}"')|
    \end{itemize}

\item \textbf{Abandon the Correct Path}
    \begin{itemize}[leftmargin=4.0em, labelindent=0pt, itemsep=3pt, topsep=3pt, parsep=2pt]
        \item[\textbf{Thought:}] I'll rebuild to verify my fix.
        \item[\textbf{Tool Call:}] \lstinline|run_shell(shell_command='./gradlew assembleDebug --parallel')|
        \item[\textbf{Result:}] {\ttfamily BUILD FAILED}. The same data binding error persists. My fix didn't work. I must have been wrong. The *real* problem must be the Kotlin version warning I saw in the first log. I will ignore the data binding error and focus on fixing the build environment.
    \end{itemize}
    
\item \textbf{Descend into "Version Hell"}
    \begin{itemize}[leftmargin=4.0em, labelindent=0pt, itemsep=3pt, topsep=3pt, parsep=2pt]
        \item[\textbf{Thought:}] The project uses Kotlin `2.0.20`, which Kapt doesn't support. This must be the root cause. I will downgrade everything. I will downgrade the Kotlin version, the Compose plugin version, and the Compose BOM version to older, more stable releases.
        \item[\textbf{Tool Call:}] \lstinline|replace(file_path='.../build.gradle.kts', old_string='val kotlinVersion = "2.0.20"', new_string='val kotlinVersion = "1.9.22"')|
        \item[\textbf{Tool Call:}] \lstinline|replace(file_path='.../app/build.gradle.kts', old_string='...compose") version "2.0.20"', new_string='...compose") version "1.5.10"')|
    \end{itemize}
    
\item \textbf{Create New Errors}
    \begin{itemize}[leftmargin=4.0em, labelindent=0pt, itemsep=3pt, topsep=3pt, parsep=2pt]
        \item[\textbf{Thought:}] I have downgraded the versions. I will try building again.
        \item[\textbf{Tool Call:}] \lstinline|run_shell(shell_command='./gradlew assembleDebug --parallel')|
        \item[\textbf{Result:}] {\ttfamily BUILD FAILED}. A new error appears: {\ttfamily \lstinline|Plugin [id: 'org.jetbrains.kotlin.plugin.compose', version: '1.5.10']| was not found}. My attempt to fix the versions has made things worse.
    \end{itemize}

\item \textbf{Failure}
    \begin{itemize}[leftmargin=4.0em, labelindent=0pt, itemsep=3pt, topsep=3pt, parsep=2pt]
        \item[\textbf{Thought:}] I am stuck in a loop. Every change I make to the versioning either fails or creates a new problem. My attempts to fix the build environment have been fruitless. I have exhausted my strategies.
        \item[\textbf{Result:}] The agent terminates after reaching its maximum number of attempts. It never solved the problem because it abandoned the correct debugging path and got lost chasing a red herring.
    \end{itemize}

\end{enumerate}
\end{failurebox}

\subsection{Case 3: Unresolved Dependency and Flawed Reasoning}
\label{app:case_study_dependency}

This case study demonstrates a scenario where the build fails due to a missing dependency. The error log contains numerous `Unresolved reference` errors, all pointing to a single missing library. While both agents correctly identify the missing dependency, the shell-based agent fails to implement the fix correctly, getting stuck in a loop of syntactical errors and flawed reasoning. In contrast, GradleFixer methodically adds the dependency and successfully resolves the build.

\subsection*{Build Error Log}
The build fails during Kotlin compilation with a large number of `Unresolved reference` errors, all located in the `MediaMetadata.kt` file. The missing references, such as `drew`, `ImageMetadataReader`, and various `Exif...Directory` classes, strongly indicate that the `com.drewnoakes:metadata-extractor` library is missing.

\begin{lstlisting}[language=bash, caption={Build Failure due to Missing Dependency}, label={lst:unresolved_dependency_error}, basicstyle=\ttfamily\scriptsize]
e: .../MediaMetadata.kt:19:12 Unresolved reference 'drew'.
e: .../MediaMetadata.kt:20:12 Unresolved reference 'drew'.
e: .../MediaMetadata.kt:215:46 Unresolved reference 'ImageMetadataReader'.
e: .../MediaMetadata.kt:219:47 Unresolved reference 'ExifIFD0Directory'.
e: .../MediaMetadata.kt:239:47 Unresolved reference 'ExifSubIFDDirectory'.
e: .../MediaMetadata.kt:256:47 Unresolved reference 'GpsDirectory'.
...
> Task :app:compileX86DebugKotlin FAILED
> Task :app:compileUniversalDebugKotlin FAILED
> Task :app:compileArm64-v8aDebugKotlin FAILED
> Task :app:compileArmeabi-v7aDebugKotlin FAILED
\end{lstlisting}

\subsubsection{GradleFixer (Successful Trajectory)}
GradleFixer correctly identifies the missing dependency from the error log and systematically adds it to the project's build configuration using the version catalog, leading to a successful build.

\begin{successbox}[Successful Trajectory: GradleFixer]
\begin{enumerate}[label=\textbf{Step \arabic*:}, wide, labelwidth=!, labelindent=0pt, topsep=8pt, itemsep=10pt]

\item \textbf{Diagnose the Missing Dependency}
    \begin{itemize}[leftmargin=4.0em, labelindent=0pt, itemsep=3pt, topsep=3pt, parsep=2pt]
        \item[\textbf{Thought:}] The build is failing with many "Unresolved reference" errors pointing to `MediaMetadata.kt`. The missing references like `ImageMetadataReader` and `ExifIFD0Directory` suggest a problem with an image metadata extraction library. I will investigate the project's dependencies to resolve this.
        \item[\textbf{Tool Call:}] \lstinline|read_file(path='Gallery/gradle/libs.versions.toml')|
    \end{itemize}

\item \textbf{Identify Dependency Management Strategy}
    \begin{itemize}[leftmargin=4.0em, labelindent=0pt, itemsep=3pt, topsep=3pt, parsep=2pt]
        \item[\textbf{Thought:}] The project uses a TOML version catalog. I've confirmed that the `metadata-extractor` library is not defined. I need to add it to `libs.versions.toml` and then include it in the app's build script.
        \item[\textbf{Tool Call:}] \lstinline|read_file(path='Gallery/app/build.gradle.kts')|
    \end{itemize}

\item \textbf{Add Dependency to Version Catalog}
    \begin{itemize}[leftmargin=4.0em, labelindent=0pt, itemsep=3pt, topsep=3pt, parsep=2pt]
        \item[\textbf{Thought:}] I will now add the `metadata-extractor` library to the `libs.versions.toml` file. First, I'll add the version to the `[versions]` section, then the library definition to the `[libraries]` section to maintain consistency.
        \item[\textbf{Tool Call:}] \lstinline|replace(file_path='Gallery/gradle/libs.versions.toml', old_string='...', new_string='...')|
        \item[\textbf{Tool Call:}] \lstinline|replace(file_path='Gallery/gradle/libs.versions.toml', old_string='...', new_string='...')|
    \end{itemize}

\item \textbf{Add Dependency to App Module}
    \begin{itemize}[leftmargin=4.0em, labelindent=0pt, itemsep=3pt, topsep=3pt, parsep=2pt]
        \item[\textbf{Thought:}] Now that the library is defined in the version catalog, I will add it as an `implementation` dependency in the `app/build.gradle.kts` file.
        \item[\textbf{Tool Call:}] \lstinline|replace(file_path='Gallery/app/build.gradle.kts', old_string='...', new_string='...')|
    \end{itemize}

\item \textbf{Verify the Fix}
    \begin{itemize}[leftmargin=4.0em, labelindent=0pt, itemsep=3pt, topsep=3pt, parsep=2pt]
        \item[\textbf{Thought:}] The dependency has been added to the project configuration. I will now run the build to confirm that the issue is resolved.
        \item[\textbf{Tool Call:}] \lstinline|run_build()|
        \item[\textbf{Result:}] {\ttfamily BUILD SUCCESSFUL}. The missing dependency was the root cause. The task is complete.
    \end{itemize}

\end{enumerate}
\end{successbox}

\subsubsection{Gemini-CLI with Shell (Failed Trajectory)}
The shell-based agent also correctly identifies the missing dependency and even manages to fix the build temporarily. However, it demonstrates flawed reasoning by incorrectly concluding its own correct fix was a temporary workaround, reverting it, and then getting stuck in an unrecoverable loop of file access errors and syntax mistakes.

\begin{failurebox}[Failed Trajectory: Gemini-CLI with Shell]
\begin{enumerate}[label=\textbf{Step \arabic*:}, wide, labelwidth=!, labelindent=0pt, topsep=8pt, itemsep=10pt]

\item \textbf{Initial Diagnosis and Correct Fix}
    \begin{itemize}[leftmargin=4.0em, labelindent=0pt, itemsep=3pt, topsep=3pt, parsep=2pt]
        \item[\textbf{Thought:}] The errors in `MediaMetadata.kt` indicate a missing dependency. The code uses classes from the `com.drew...` package, so I need to add the `metadata-extractor` library. I will add it to `libs.versions.toml` and `app/build.gradle.kts`.
        \item[\textbf{Tool Call:}] (Agent successfully adds the dependency to both files.)
    \end{itemize}

\item \textbf{Flawed Reasoning and Reversal}
    \begin{itemize}[leftmargin=4.0em, labelindent=0pt, itemsep=3pt, topsep=3pt, parsep=2pt]
        \item[\textbf{Thought:}] I've added the `metadata-extractor` dependency to fix the issue. I'll now remove the artifacts made to the build files.
        \item[\textbf{Tool Call:}] (Agent reverts its own correct changes from both `app/build.gradle.kts` and `gradle/libs.versions.toml`.)
    \end{itemize}
    
\item \textbf{Unnecessary Cache Cleaning}
    \begin{itemize}[leftmargin=4.0em, labelindent=0pt, itemsep=3pt, topsep=3pt, parsep=2pt]
        \item[\textbf{Thought:}] I reverted the changes. Now, I will try to build again. (Build fails with the original error). The build failed again. This often indicates a corrupted cache. I'll clean the build cache and try building again.
        \item[\textbf{Tool Call:}] \lstinline|run_shell(shell_command='./gradlew cleanBuildCache')|
        \item[\textbf{Result:}] Command fails as `cleanBuildCache` is not a valid task. The agent then tries `clean`.
    \end{itemize}

\item \textbf{Rediscovery and Repeated Failures}
    \begin{itemize}[leftmargin=4.0em, labelindent=0pt, itemsep=3pt, topsep=3pt, parsep=2pt]
        \item[\textbf{Thought:}] It appears I mistakenly reverted the fix that should have resolved the build error. My apologies. The "Unresolved reference" errors indicate that the `metadata-extractor` dependency is indeed required. I will now re-apply the fix.
        \item[\textbf{Tool Call:}] (Agent now enters a long loop of failures, repeatedly attempting to modify the Gradle files but failing due to incorrect syntax for referencing version catalog aliases with hyphens, e.g., using `libs.metadata.extractor` instead of the correct `libs.metadataExtractor`.)
    \end{itemize}

\item \textbf{Failure}
    \begin{itemize}[leftmargin=4.0em, labelindent=0pt, itemsep=3pt, topsep=3pt, parsep=2pt]
        \item[\textbf{Thought:}] My apologies for the repeated mistakes. It seems I'm stuck in a loop. I will try to fix the issue by first re-reading the `gradle/libs.versions.toml` file to confirm the correct dependency alias.
        \item[\textbf{Result:}] The agent continues to struggle with the correct syntax and file paths, unable to break out of its error loop. It eventually terminates after reaching its maximum number of attempts, having failed to re-apply the fix it had already discovered once.
    \end{itemize}

\end{enumerate}
\end{failurebox}

\clearpage
\newpage

\begin{strip}
    \centering
    \captionof{table}{List of the 43 projects included in the AndroidBuildBench benchmark, with key statistics for each project. LOC refers to lines of code.}
    \label{tab:project_list_appendix_full}
    \small % or \tiny
    \sisetup{group-separator={,}}
    % The tabular environment is NOT a float, so it's perfectly fine here.
    \begin{tabular}{@{}l S[table-format=5.0] S[table-format=4.0] S[table-format=5.0] S[table-format=4.0] S[table-format=6.0] l@{}}
      \toprule
      \textbf{Project} & {\textbf{Stars}} & {\textbf{Closed PRs}} & {\textbf{Commits}} & {\textbf{Files}} & {\textbf{LOC}} & {\textbf{Last Commit}} \\
      \midrule
        android/nowinandroid & 19300 & 1216 & 2897 & 681 & 614619 & 12 hours ago \\
        android/sunflower & 17800 & 679 & 577 & 162 & 35207 & 1 year ago \\
        android/uamp & 13200 & 142 & 456 & 181 & 13597 & 3 weeks ago \\
        thunderbird/thunderbird-android & 12300 & 4197 & 16637 & 6393 & 191651 & 5 hours ago \\
        Shabinder/SpotiFlyer & 10900 & 210 & 606 & 491 & 78745 & 9 months ago \\
        AppIntro/AppIntro & 10600 & 592 & 1118 & 215 & 131060 & 3 weeks ago \\
        T8RIN/ImageToolbox & 9000 & 1252 & 8459 & 2672 & 276038 & 2 hours ago \\
        recloudstream/cloudstream & 7800 & 869 & 2928 & 1192 & 173965 & 2 days ago \\
        aniyomiorg/aniyomi & 6200 & 487 & 8102 & 1988 & 14332 & 3 weeks ago \\
        LibChecker/LibChecker & 5700 & 1397 & 2442 & 767 & 132695 & 4 days ago \\
        Tapadoo/Alerter & 5500 & 101 & 456 & 138 & 185834 & 4 years ago \\
        nextcloud/android & 4700 & 7082 & 28359 & 2710 & 188193 & 14 hours ago \\
        Aliucord/Aliucord & 4200 & 227 & 522 & 269 & 73082 & 1 week ago \\
        owncloud/android & 4000 & 2116 & 13760 & 2710 & 188193 & 3 days ago \\
        jarnedemeulemeester/findroid & 3300 & 475 & 1507 & 695 & 164493 & 2 weeks ago \\
        wikimedia/apps-android-wikipedia & 2700 & 5759 & 20931 & 2296 & 118073 & 7 hours ago \\
        spacecowboy/Feeder & 2200 & 430 & 3806 & 913 & 352096 & 2 days ago \\
        kylecorry31/Trail-Sense & 2000 & 1531 & 10213 & 4143 & 10588 & 1 day ago \\
        IacobIonut01/Gallery & 1800 & 244 & 751 & 544 & 325656 & 2 months ago \\
        flipperdevices/Flipper-Android-App & 1800 & 892 & 1019 & 3114 & 32046 & 2 weeks ago \\
        keymapperorg/KeyMapper & 1700 & 196 & 4505 & 1186 & 151367 & 2 weeks ago \\
        Mahmud0808/ColorBlendr & 1500 & 119 & 783 & 555 & 86787 & 2 days ago \\
        TrianguloY/URLCheck & 1500 & 202 & 989 & 409 & 94699 & 4 days ago \\
        MMRLApp/MMRL & 1400 & 233 & 2754 & 858 & 88845 & 2 weeks ago \\
        Pool-Of-Tears/Myne & 1300 & 171 & 242 & 306 & 123231 & 1 week ago \\
        GetStream/whatsApp-clone-compose & 1300 & 309 & 468 & 219 & 125767 & 5 months ago \\
        avluis/Hentoid & 1200 & 209 & 9973 & 1160 & 176125 & 4 days ago \\
        LemmyNet/jerboa & 1200 & 970 & 1253 & 359 & 128027 & 3 days ago \\
        GrapheneOS/Camera & 1100 & 196 & 1482 & 205 & 28302 & 2 weeks ago \\
        AllanWang/Frost-for-Facebook & 1100 & 394 & 1263 & 1016 & 56033 & 2 years ago \\
        patzly/grocy-android & 1000 & 237 & 5028 & 1069 & 287205 & 3 days ago \\
        Stypox/dicio-android & 1000 & 101 & 1540 & 599 & 33686 & 1 month ago \\
        nextcloud/notes-android & 1000 & 1542 & 4699 & 715 & 94640 & 14 hours ago \\        NordicSemiconductor/Android-DFU-Library & 813 & 133 & 870 & 459 & 59219 & 1 month ago \\
        getodk/collect & 736 & 3616 & 17032 & 2395 & 59552 & 3 days ago \\
        Pool-Of-Tears/GreenStash & 702 & 115 & 253 & 287 & 138502 & 1 month ago \\
        d4rken-org/capod & 695 & 150 & 667 & 1302 & 108508 & 1 month ago \\
        leonlatsch/Photok & 687 & 251 & 1189 & 409 & 47288 & 1 day ago \\
        DroidKaigi/conference-app-2021 & 640 & 479 & 1999 & 883 & 47775 & 4 years ago \\
        kasem-sm/SlimeKT & 606 & 176 & 111 & 547 & 102708 & 2 years ago \\
        android/socialite & 576 & 101 & 410 & 204 & 43609 & 5 days ago \\
        oxygen-updater/oxygen-updater & 565 & 175 & 1144 & 392 & 41489 & 9 hours ago \\
        you-apps/ClockYou & 554 & 170 & 894 & 274 & 24116 & 1 month ago \\
      \bottomrule
    \end{tabular}
\end{strip}

\section{Curated Projects in AndroidBuildBench}
\label{app:project_list}
AndroidBuildBench is curated from 43 diverse, popular, and actively maintained open-source Android projects from GitHub. The full list of projects, identified by their username/repository, is provided in Table \ref{tab:project_list_appendix_full}.

\end{document}